\documentclass[12pt]{article}
\usepackage{pdflscape}
\usepackage{tikz,natbib,xfrac,amsmath,amsthm,booktabs,multirow,subfigure,anysize,graphicx,float,enumitem,eurosym}
\definecolor{dark-red}{rgb}{0.4,0.15,0.15}
\definecolor{dark-blue}{rgb}{0.15,0.15,0.4}
\definecolor{medium-blue}{rgb}{0,0,0.5}
\definecolor{ChadBlue}{rgb}{.1,.1,.5}
\definecolor{ChadDarkBlue}{rgb}{.1,0,.2}
\definecolor{ChadBlue}{rgb}{.1,.1,.5}
\definecolor{ChadRoyal}{rgb}{.2,.2,.8}
\definecolor{ChadGreen}{rgb}{0,.4,0}    
\definecolor{ChadRed}{rgb}{.5,0,.5}  
\usepackage[T1,hyphens]{url}
\usepackage[colorlinks,urlcolor=medium-blue]{hyperref}
\hypersetup{
	colorlinks, linkcolor={ChadRed},
	citecolor={ChadBlue}}
\usepackage[margin=1in]{geometry}
\usepackage{xcolor}
\usepackage{bm}

\usepackage[font=small,labelfont=bf,justification=justified,singlelinecheck=false]{caption}
\usepackage{amssymb}
\usepackage{amsfonts}
\usepackage{amsmath}
\usepackage{setspace}  
\usepackage[all]{hypcap}        

\graphicspath{{figures/}}
\usepackage[english]{babel}
\usepackage{longtable}
\usepackage[singlelinecheck=off]{caption}
\usepackage[singlelinecheck=false]{caption}
\usepackage{array}
\usepackage{float}

\usepackage [autostyle, english = american]{csquotes}
\MakeOuterQuote{"}
\usepackage{chngcntr}
\usepackage{rotating, graphicx}
\usepackage{epstopdf}
\usepackage{threeparttable}
\usepackage{sectsty}
\sectionfont{\large}
\usepackage[titletoc,title]{appendix}

\newcommand{\beginappendix}{%
	\setcounter{table}{0}
	\renewcommand{\thetable}{A\arabic{table}}%
	\setcounter{figure}{0}
	\renewcommand{\thefigure}{A\arabic{figure}}%
}
\usepackage{longtable}
\usepackage{booktabs}
\usepackage{array}
\newcolumntype{R}{>{\raggedleft\arraybackslash}p{2.5cm}}
\newcolumntype{Q}{>{\raggedright\arraybackslash}p{6cm}}
\usepackage{tabularx}
\usepackage{dcolumn}
\usepackage[T1]{fontenc}
\usepackage{chronosys}

\title{How exporters neutralized an increase in tariffs\thanks{I thank Elena Cifuentes, Raúl Mínguez, Rafael Pico, Juán Manuel Sánchez, Cristina Vega, Albert Vinaixa and participants at ETSG 2022 Groningen and the 47th Simposio de An\'{a}lisis Econ\'{o}mico for their valuable comments and suggestions. I acknowledge the Department of Customs and Excise of the Spanish Tax Agency (AEAT) and the Spanish Chamber of Commerce for providing customs data. This research was conducted as part of the Project PID2021-122133NB-I00 financed by MCIN/AEI/10.13039/501100011033/FEDER, EU. I also gratefully acknowledge the financial support from the Basque Government Department of Education (IT1429-22). Part of the research presented in this paper was done while I visited the Department of Economics and the Real Colegio Complutense at Harvard University. I gratefully acknowledge their hospitality. I also thank the Basque Government's Ikermugikortasuna 2022 program for financing the research stay.} }

\author{\large {Asier Minondo}\thanks{Deusto Business School, University of Deusto, Camino de Mundaiz 50, 20012 Donostia - San Sebasti\'{a}n (Spain). Email: \href{mailto:aminondo@deusto.es}{aminondo@deusto.es}}  \\}

\date{ {This version:} \today \\  }

\begin{document}
\maketitle

\begin{abstract}
I use the unanticipated and large additional tariffs the US imposed on European Union products due to the Airbus-Boeing conflict to analyze how exporters reacted to a change in trade policy. Using firm-level data for Spain and applying a difference-in-differences methodology, I show that the export revenue in the US of the firms affected by the tariff hike did not significantly decrease relative to the one of other Spanish exporters to the US. I show that Spanish exporters were able to neutralize the increase in tariffs by substituting Spanish products with products originated in countries unaffected by tariffs and shifting to varieties not affected by tariffs. My results show that tariff avoidance is another margin exporters can use to counteract the effects of a tariff hike.
\end{abstract}

\begin{flushleft}
	\textbf{JEL}: F10, F14\
\end{flushleft}
\textbf{Keywords}: tariffs, tariff avoidance, exporting firms, Spain, US, Airbus-Boeing conflict.

\newpage 
\onehalfspacing

\section{Introduction}
\label{sec:introduction}

With the arrival of Donald Trump to the US presidency, tariffs became a major policy tool of the US administration. As recorded by the Peterson Institute's Trump's Trade War Timeline, the Trump administration opened six major trade wars, which affected all its main trade partners.\footnote{\url {https://www.piie.com/blogs/trade-investment-policy-watch/trump-trade-war-china-date-guide}.} The US also withdrew from the Trans-Pacific Partnership and forced a renegotiation of the North American Free Trade Agreement seeking to restrict its free-trade provisions. The regression from trade integration was not a trend exclusive to the US \citep{colantone2021backlash}. For example, in January 2020, the United Kingdom disengaged from the European Union (EU), culminating the biggest reversal in deep economic integration in recent history \citep{dhingra2022brexit}. 


The shift toward more aggressive and unilateral trade policies has raised interest in understanding how exporters adjust to a more hostile environment. However, there is scant evidence on how economies and firms in practice respond to tariffs \citep{cavallo2021passthrough}. This paper analyzes how Spanish exporters adjusted to the tariffs imposed by the Trump Administration because of the Airbus-Boeing conflict. In October 2019, the US introduced an additional 25\% tariff on 64 Spanish products, which represented almost 5\% of the total Spanish merchandise exports to the US in 2018.\footnote{The US was the top-6 destination of Spanish merchandise exports, accounting for 4.4\% of its total exports in 2018.}  Furthermore, the importance of the US market for the Spanish exporters affected by the tariff was high: 32\%. A key feature of this trade conflict is that the US imposed tariffs on many products that had no relationship with the aircraft industry. For example, in the case of Spain, the products most affected by the additional tariff were olive oil, wine, cheese, and olives. Hence, exporters of those products could not anticipate the imposition of a tariff. This creates a suitable environment to identify how exporters adjust to a change in trade policy.

Using firm-level monthly exports data to the US and a difference-in-differences specification, I find that Spanish exporters affected by the increase in tariffs experienced an insignificant decrease in export revenue in the US relative to other Spanish exporters to the US. Spanish exporters affected by the additional tariff did not have either a larger probability of exiting the US market than the rest of the exporters to the US. 


How were Spanish exporters to the US able to counteract the increase in tariffs? They used two major strategies to \emph{avoid} tariffs. The first one was to substitute a product originated in Spain by a product originated in a country not targeted by tariffs. For example, Spanish olive oil exporters to the US increased their imports of olive oil from Tunisia and Portugal, countries not affected by the US tariff, bottled it in Spain, and shipped it to the US with a label declaring that the origin of the product was Tunisia or Portugal. Spanish Customs recorded these shipments as Spanish exports to the US. However, as was the goal of Spanish exporters, US Customs considered that the bottling of the olive oil in Spain had not introduced a substantial change in the product, and thus recorded the shipments from Spain as imports from Tunisia or Portugal, with no tariffs imposed. Due to this recording discrepancy, the value of olive oil imports from Spain recorded by US customs accounted for less than 50\% of the mirror export value recorded by Spanish customs. I provide evidence showing that this gap was not the result of tariff evasion, but the outcome of a tariff-avoidance strategy followed by Spanish exporters.

The second strategy, used by wine and cheese producers, was to shift exports from tariff-targeted varieties to similar non-targeted varieties. For example, wines with an alcohol strength by volume of 14\% or less were subject to an additional tariff, whereas wines with a higher alcohol strength were not. The alcoholic strength of red wines, the most important variety exported by Spain to the US, is in the 13-15 range. Therefore, it was feasible for exporters of red wine with an alcoholic strength equal or lower than 14\% to shift their exports to higher alcoholic strength varieties to avoid the tariff. Regarding cheese, some varieties of cheese made from sheep's milk, the most important cheese variety exported by Spain to the US, were targeted by a tariff, whereas other similar ones were not. This enabled Spanish exporters to shift from the targeted to the non-targeted variety and avoid the negative impact of the tariff. I show that there was an actual change in the exported varieties and not a product reclassification to evade tariffs. I document that targeted varieties that were ``far'' in product characteristics from the non-targeted ones (e.g., white wine and cheese made from cow's milk) experienced a much larger decrease in exports than the ``close'' ones.

The strategies implemented by Spanish firms during the Airbus-Boeing conflict reveal that exporters have another margin to adjust to changes in trade policy: trade avoidance. Like the trade evasion margin highlighted by \cite{demir2020benfordslaw}, trade avoidance enables exporters, at least the ones specialized in primary and the manufacturing of primary products, to respond swiftly to changes in trade policy. This paper shows that this margin can be powerful enough to neutralize the effect of an increase in tariffs. 

This paper contributes to four strands of literature. Firstly, this work relates to the margins exporters have to adjust to changes in trade policy. Models of international trade with export fixed costs and firm heterogeneity in productivity show that some exporters will reduce their sales (the intensive margin) and others will exit export markets (the extensive margin) in response to a raise in tariffs \citep{melitz2003impact}. Firms exporting multiple products (the product margin) will concentrate their sales in their most productive, or core, product in response to a raise in trade costs \citep{eckel2010multi,bernard2011multiproduct}. Models with strategic pricing (the mark-up margin) predict that firms will reduce their mark-ups in response to a raise in tariffs \citep{atkeson2008pricing}. \cite{ludema2016passthrough} showed that an increase in tariffs lead exporters to reduce the quality of products (the quality margin). Finally, \cite{demir2020benfordslaw} showed that exporters in Turkey increased tariff evasion after taxes were introduced on some types of foreign transactions (the tariff-evasion margin). I show that tariff avoidance is another margin exporters can use to mute the negative effects of a tariff increase.

Secondly, the paper contributes to the empirical literature that analyzes how exporters react to changes in trade policy. \cite{fitzgerald2018shocks} showed that exporters are more responsive to changes in tariffs than changes in the real exchange rate. \cite{albornoz2021tariffhikes} concluded that fewer Argentinian firms exported to the US after an unanticipated tariff was imposed on their products. They also showed that surviving exporters began to sell new products in the US to counteract the reduction in revenue from the tariff-targeted products. \cite{jiao2022impacts} found that exports of Chinese firms to the US dropped significantly after the US imposed tariffs on their products. This drop was only partially compensated with higher sales to the EU. This paper contributes to this literature showing that exporters can mute the negative effects of tariffs exporting products originated in other countries and shifting to varieties not targeted by tariffs.

Thirdly, this paper is related to the tariff \emph{evasion} literature. In a seminal article,  \cite{bhagwati1964underinvoicing} compared the value of exports reported by the exporter country and the value of mirror imports reported by the importer country to evaluate the magnitude of tariff evasion in Turkey. This methodology was recovered and refined by \cite{fisman2004tax}, who showed that misreporting in Chinese imports from Hong-Kong was positively correlated with tariff rates. Misreporting happened mainly because importers misclassified high-tariff products as similar low-tariff products. \cite{javorcik2008differentiated} found that tariff evasion is more prevalent in differentiated than homogeneous products, because it is more difficult to identify a price misreporting in the former than in the latter. This study contributes to this literature showing that the positive correlation between the trade gap between exports and mirror imports and tariffs may also be explained by a tariff-avoidance strategy rather than a tariff-evasion one.

Fourthly, this paper contributes to the tariff avoidance literature. One of the best known examples of tariff avoidance is the one related to the so-called chicken tax. In the early 1960s, Germany imposed a tax on frozen chicken imported from the US. In retaliation, the US imposed a 25\% global tariff on light trucks, with the goal of making the import of German Volkswagen pick-ups, which were very popular in the US at that time, less attractive. The tariff remained in force even when disputes between Germany and the US over chicken disappeared. Paradoxically, the chicken tax ended up hurting US car manufacturers, which used very awkward procedures to avoid the tariff when importing the vans produced in their production facilities abroad into the US.\footnote{See, for example, Wall Street Journal ``To Outfox the Chicken Tax, Ford Strips Its Own Vans'' published on September 23, 2009. Available at \url{https://www.wsj.com/articles/SB125357990638429655\#pr}.} There is also literature showing that firms shift production across export platforms to avoid the effect of tariffs \citep{flaaen2020production}. This paper contributes to this literature showing that exporters can implement tariff-avoidance strategies very rapidly in response to a change in trade policy. 

The paper is organized as follows. Section~\ref{sec:airbus_boeing_tariff} presents the timeline of the Airbus-Boeing conflict, highlighting the lengthy process that culminated in the imposition of tariffs by the US. It also identifies the Spanish products that were more affected by tariffs. Section~\ref{sec:impact_on_trade} uses econometric analyses to estimate the impact of tariffs on exports at the product and firm level. Section~\ref{sec:tariff_avoidance_strategies} describes the tariff-avoidance strategies followed by Spanish exporters, and the final section concludes.

\section{The Airbus-Boeing conflict}
\label{sec:airbus_boeing_tariff}

On October 6, 2004, the US requested consultations in the World Trade Organization (WTO) with France, Germany, Spain, United Kingdom, and the EU related to potential subsidies received by Airbus.\footnote{The code granted by the WTO to this case was DS316. All the documents related to this dispute can be found at \url{https://www.wto.org/english/tratop_e/dispu_e/cases_e/ds316_e.htm}.} The main complaint was that Airbus had received preferential loans to design and develop new aircraft models. Allegedly, this had enabled Airbus to market new products faster and at a lower cost than in a situation in which loans had been granted on market terms. After a lengthy process, in June 2011, the WTO Dispute Settlement Body ruled that Airbus had received subsidies that were inconsistent with the WTO rules and requested the EU to remove them. In December 2011, the EU asserted that it had removed the illegal subsidies. However, the US argued that the illegal subsidies remained in force and requested the formation of a compliance panel. On May 28, 2018, the Dispute Settlement Body ruled that the EU had not removed the illegal subsidies and allowed the US to take retaliatory measures. In October 2019, the WTO Arbitrator concluded that the US could apply countermeasures on \$7.5 billion of imports from the EU, the largest retaliatory measure granted in the WTO history at that time.\footnote{In a parallel case that began the very same day the US requested consultations in the WTO, the EU requested consultations with the US because it suspected that Boeing had received illegal subsidies from some US states. After an exceptionally long process, in March 2019, the WTO ruled that the US had not withdrawn the illegal subsidies, that those subsidies were damaging EU interests, and allowed the EU to take retaliatory measures against the US. These measures, which amounted to \$4 billion of imports from the US, were implemented in November 2020.}

\begin{figure}[t]
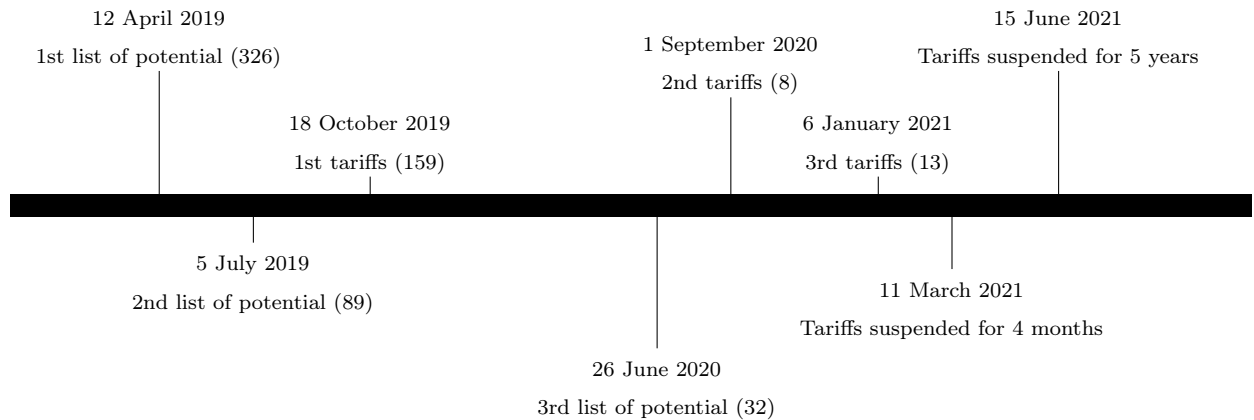

	\caption{Timeline of the Airbus-Boeing conflict since April 2019}
	\label{fig:timeline}
\startchronology[startyear=2019,stopyear=2022,startdate=FALSE,stopdate=FALSE,arrow=FALSE]\setupchronoevent{textstyle=\scriptsize,datestyle=\scriptsize}
\chronoevent[markdepth=-60pt]{12/4/2019}{1st list of potential (326)}
\chronoevent[markdepth=10pt]{5/7/2019}{2nd list of potential (89)}
\chronoevent[markdepth=-20pt]{18/10/2019}{1st tariffs (159)}
\chronoevent[markdepth=50pt]{26/6/2020}{3rd list of potential (32)}
\chronoevent[markdepth=-50pt]{1/9/2020}{2nd tariffs (8)}

\chronoevent[markdepth=-20pt]{6/1/2021}{3rd tariffs (13)}

\chronoevent[markdepth=20pt]{11/3/2021}{Tariffs suspended for 4 months}

\chronoevent[markdepth=-60pt]{15/6/2021}{Tariffs suspended for 5 years}
\stopchronology
\footnotesize Note: The number in parentheses corresponds to the number of products affected by the measure.
\end{figure}

On April 12, 2019, the US government presented a preliminary list of 326 EU products that could be targeted by an additional tariff. Figure~\ref{fig:timeline} presents the timeline of the conflict since that date. As expected, aircrafts were included in the list. However, most of the products included in the list were not related to the aircraft industry (e.g., salmon fillets, yogurt, handbags, or printed books). In some products, all EU countries were potential targets of the additional tariff. However, in other products, only some EU countries were affected. The US specifically targeted the EU countries where Airbus production facilities were located and which have granted subsidies to the aircraft company: France, Germany, Spain, and the United Kingdom. On the 5th of July 2019, the list of products potentially targeted by tariffs was enlarged with 89 new goods.

\catcode`\@=11
\def\chron@selectmonth#1{\ifcase#1\or January\or February\or
	March\or April\or May\or June\or
	July\or August\or September\or
	October\or November\or December\fi}

On October 9, 2019, the US announced the products that were affected by an additional tariff. Specifically, 159 out of 405 potential products were affected by the additional tariff (39\%). For all products, except for the aircraft industry, the additional tariff was set at 25\%.\footnote{The additional tariff for aircraft-related products was set at 10\%. However, the tariff was increased to 15\% in February 2020.} These tariffs came into force on October 18, 2019. In June 2020, 30 more products were added to the potential list of targeted products. In September 2020 and January 2021, additional tariffs were applied on 8 and 13 more products, respectively.\footnote{In March 2020, the tariff was lifted from one product (prune juice) and applied to another product (butchers' or kitchen chopping or mincing knives).} 138 out of the 180 products targeted by tariffs were primary products or manufactures of primary products, such as cheese, fruits and nuts, olive oil, olives, wine, and liquors. The remaining 42 products were manufactures, such as aircrafts, printed pictures, or sweaters. Finally, in March 2021, the US and the EU agreed to suspend the tariffs for 4 months. The temporary suspension was widened to a 5-year period in June 2021.\footnote{The US government reached a similar agreement with the UK, which had already left the EU's single market and customs union at the beginning of January 2021.}

\begin{figure}[t]
	\begin{center}
		\caption{Distribution of Spanish exports to the US by tariff-targeted products (based on Spanish exports to the US in 2018}
		\label{fig:tree_map}
		\includegraphics[scale=0.4]{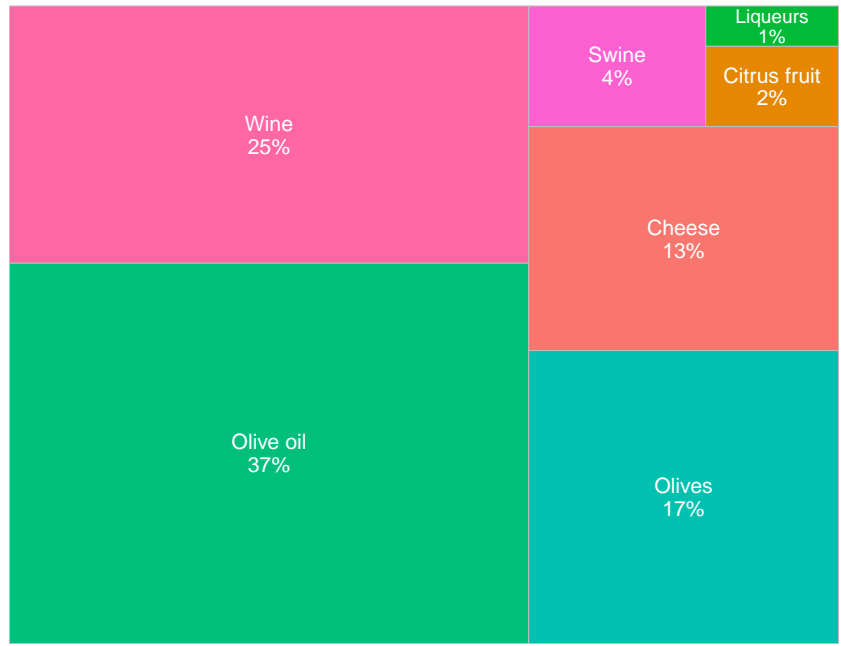}
	\end{center}
	\footnotesize Source: author's own elaboration based on US Census trade data.
\end{figure}

Regarding Spain, 277 products were potentially subject to an additional tariff. Among them, 64 products were targeted by an additional tariff in October 2019.\footnote{No Spanish product was added to the products targeted by tariffs after October 2019.} The Spanish exports to the US that were targeted by tariffs amounted to 5\% of all Spanish exports to the US in 2018. Figure~\ref{fig:tree_map} shows that the exports subjected to tariffs were concentrated in four products: olive oil, which accounted for 37\% of all Spanish exports targeted by tariffs in 2018, wine (25\%), olives (17\%), and cheese (13\%). The remaining tariff-targeted products only accounted for 8\% of tariff-targeted exports.\footnote{None of the aircraft products exported by Spain to the US were targeted by tariffs.}

Except for aircraft-related products, the US administration did not explain the criteria it had followed to select the products that could be affected by tariffs. Therefore, except for the aircraft manufacturers, the inclusion on the list was unanticipated by EU exporters. To confirm this, following \cite{albornoz2021tariffhikes}, I estimate the following equation:

\begin{equation}
\label{eq:selection_into_potential}
y_{k}=\beta_{0}+\beta_{1}Tariff_{k}+\beta_{2}M_{k}+\beta_{3}\Delta M_{k}+\epsilon_{k} 
\end{equation}

where $y_{k}$ is an indicator variable that turns one if product $k$ was included in the list of products that could potentially be targeted by tariffs, $\beta_{0}$ is a constant, $Tariff_{k}$ is the duty imposed by the US on EU product $k$ before the Airbus-Boeing conflict, $M_{k}$ is the value of US imports of product $k$ from the EU, and $\Delta M_{k}$ is the growth in imports from the EU between 2015-2016 and 2017-2018. Products are defined at the HTS 8-digit level, because tariffs, except for aircrafts, were imposed on that level.\footnote{In a few products, the potential tariff only affected some EU countries. I exclude these products from the sample.} Column~1 of Table~\ref{tab:select_potential} shows that products that had an initial higher tariff were more likely to be included in the potential list. However, this effect becomes statistically insignificant once I control for other confounding variables. The level and growth of imports do not have a significant effect on the probability to be included in the list of products targeted by tariffs.

As explained above, in virtually all targeted products the tariff duty increased by 25 percentage points. The lack of variation in the additional tariff across products and industries indicates that industry-specific pressure groups had little influence in the determination of tariffs \citep{fajgelbaum2020return}.


\begin{table}[!htbp] \centering 
  \caption{Selection of products that could potentially be targeted by tariffs} 
  \label{tab:select_potential} 
\small 
\begin{tabular}{@{\extracolsep{2.7pt}}lccccc} 
\\[-1.8ex]\hline 
\hline \\[-1.8ex] 
\\[-1.8ex] & (1) & (2) & (3) & (4) & (5)\\ 
\hline \\[-1.8ex] 
 Tariff rate & 0.110$^{b}$ & 0.066 & 0.066 & 0.065 & $-$0.020 \\ 
  & (0.044) & (0.049) & (0.049) & (0.049) & (0.049) \\ 
  Imports EU 2018 & 1.112 & 5.471 & 5.471 & 6.755 & 5.094 \\ 
  & (3.231) & (5.504) & (5.504) & (6.063) & (5.612) \\ 
  Imports growth EU &  &  & 0.000 & 0.000 & $-$0.002 \\ 
  &  &  & (0.002) & (0.002) & (0.002) \\ 
  Imports non-EU 2018 &  &  &  & $-$0.776 & 0.910 \\ 
  &  &  &  & (1.553) & (1.688) \\ 
  Imports growth non-EU &  &  &  & $-$0.001 & $-$0.002 \\ 
  &  &  &  & (0.003) & (0.003) \\ 
 \hline \\[-1.8ex] 
Conditional on growth sample & \textup No & \textup Yes & \textup Yes & \textup Yes & \textup Yes \\ 
HS-4 digits fixed effects & \textup No & \textup No & \textup No & \textup No & \textup Yes \\ 
Observations & 8,926 & 8,247 & 8,247 & 8,247 & 8,247 \\ 
Adjusted R$^{2}$ & 0.001 & 0.000 & 0.000 & 0.000 & 0.371 \\ 
\hline 
\hline \\[-1.8ex] 
\end{tabular}
\centerline{\begin{minipage}{0.95\textwidth}~\
\footnotesize{Note: Standard errors in parentheses. b: Statistically significant at 5\%.
Imports EU and Imports non-EU refer to the year 2018. Import growth is computed as the log
change in imports between 2015-2016 and 2017-2018. Columns (1) and (2) differ in that column (2)
conditions on the sample of columns (3) to (5), that is, product observations for which growth rate 
can be computed.} \end{minipage}} 
\end{table}

Once the first list of potential tariffs was announced, firms could determine whether they could be affected by an increase in tariffs and take precautionary measures. However, as will be shown in the next section, there was no significant change in the behavior of exporters that were finally targeted by tariffs during the period that elapses between the publication of the list of potentially-affected products and the announcement of products that were finally targeted by tariffs. This result suggests that potentially-targeted exporters did not take precautionary measures, such as anticipating sales and building stocks in the US, to counteract the potential raise in tariffs \citep{alessandria2019taking,khan2021does}.


\section{The impact of tariffs on Spanish exports to the US}
\label{sec:impact_on_trade}

This section analyzes the impact of tariffs on Spanish exports to the US. Firstly, I explore the impact of tariffs at the product level using import data from the US Census. Next, I analyze the impact of tariffs on Spanish exporters revenue and survival in the US market using data from Spanish customs. As will be seen, the impact of tariffs was different when analyzed at the product or exporter level.

\subsection{Product-level analyses}

I use data on US monthly imports from Spain from the US Census to compare the evolution of products that were targeted by a tariff with that of those that were not.\footnote{To obtain a cleaner comparison, I excluded the products that had been included in the list of products that potentially could be targeted by a tariff but, finally, no tariff was imposed.} I calculate a 12-month rolling sum of imports to avoid the noise introduced by monthly data. For example, the value corresponding to January 2019 is the sum of the value of imports in that month and the previous 11 months. The value of imports in September 2019, one month before the introduction of tariffs, is set at 100.

\begin{figure}[htbp]
	\begin{center}
		\caption{US imports from Spain: tariff vs. non-tariff products, 2019-2022 (rolling 12-month sum; September 2019=100)}
		\label{fig:esp_tar_vs_rest}
		\includegraphics[height=3.5in]{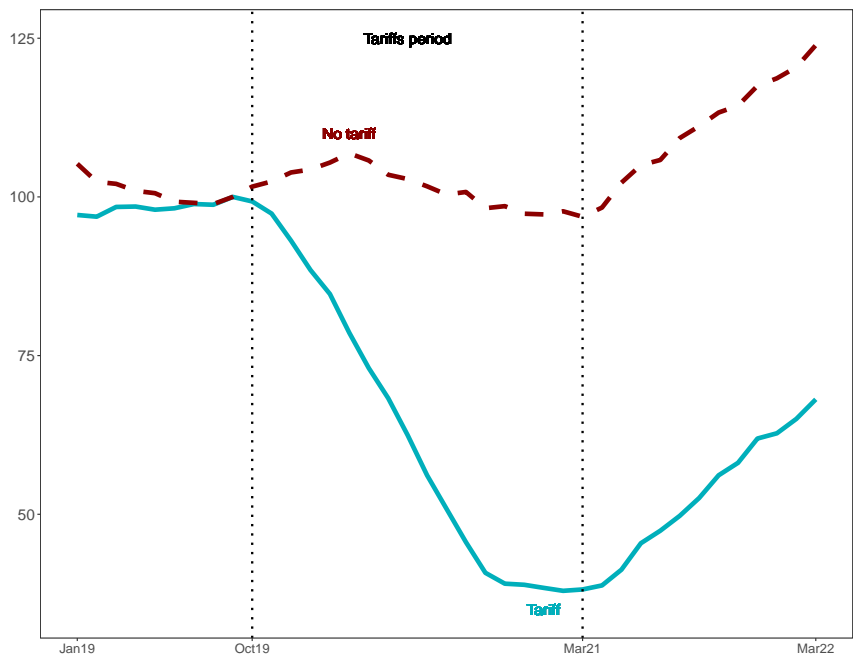}
	\end{center}
	\footnotesize Source: author's own elaboration based on the US Census database.
\end{figure}

Figure~\ref{fig:esp_tar_vs_rest} shows that tariffs led to a large decline in imports. During the period in which tariffs were in force, October 2019-March 2021, imports of products targeted by tariffs (blue solid line) decreased by 62\%. Imports of products not targeted by tariffs (red dashed line) only declined by 3\%. Imports of products targeted by tariffs began to increase once tariffs were lifted in March 2021.\footnote{Tariffs were suspended on the 4th and 11th of March for UK and EU products, respectively.} However, the import level one year after tariffs had been lifted, March 2022, was still 32\% lower than the one in the month before tariffs were imposed.

\begin{figure}[htbp]
	\begin{center}
		\caption{Impact of tariffs on the four main products targeted by tariffs, 2019-2022 (rolling 12-month sum; September 2019=100)}
		\label{fig:m_usa_4products}
		\includegraphics[height=3.5in]{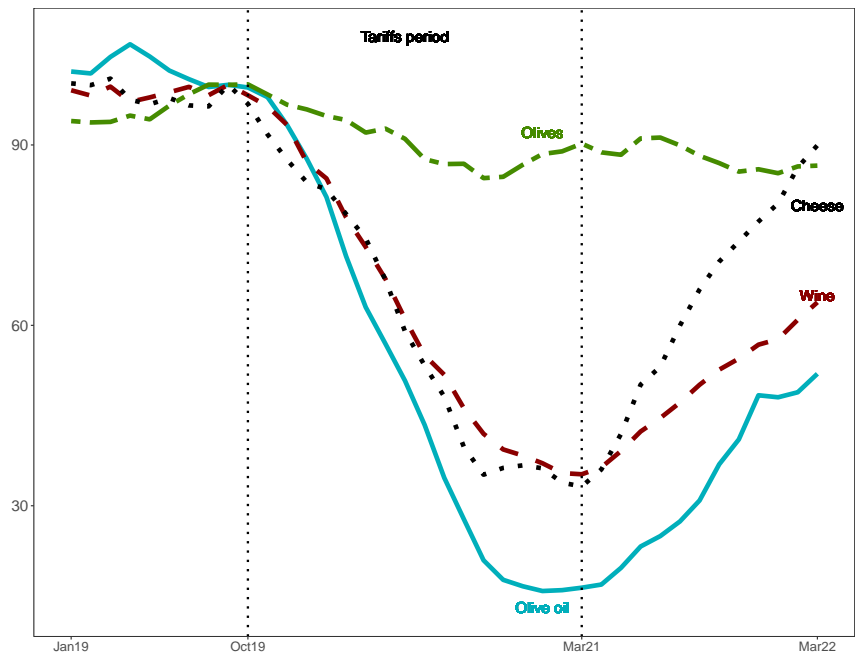}
	\end{center}
	\footnotesize Source: author's own elaboration based on the US Census database.
\end{figure}

Figure~\ref{fig:m_usa_4products} presents the evolution of imports in the four main Spanish products, in terms of value, targeted by tariffs: olive oil, cheese, wine, and olives.\footnote{For each product category, I only compute the imports of the varieties that were hit by a tariff.} The largest decrease happened in olive oil, where imports dropped by 84\% during the tariff period. Decreases were also large for wine and cheese: 65\% and 67\% respectively. The lowest decrease happened in olives: 10\%. 


To confirm the visual appreciation from Figure~\ref{fig:esp_tar_vs_rest}, I estimate the following difference-in-differences specification:

\begin{equation}
	\label{eq:product_analysis1}
	y_{ktm}=exp[\beta(Tariff_{k}*Post_{tm})+\gamma_{km}+\gamma_{t}+\epsilon_{ktm}]
\end{equation}


where $y_{ktm}$ is the import value of US imports of product $k$ at year $t$ and month $m$ from Spain. $Tariff_{k}$ turns one if the US imposed a tariff on product $k$ imported from Spain. $Post_{tm}$ turns one if the year-month combination is after October 2019.\footnote{Since tariffs were imposed in the middle of October 2019, I exclude this month's observations from the sample.} The specification includes product $\times$ month ($\gamma_{km}$) and year ($\gamma_{t}$) fixed effects. $\epsilon_{ktm}$ is the disturbance term. To exploit the granularity of data, products are defined at the 10-digit HTS classification. The $\beta$ coefficient captures whether the difference in the dependent variable between non-tariff-targeted and tariff-targeted products (first difference) changed after tariffs were imposed (second difference). I cluster standard errors at the product level. 


\begin{table}[htbp]
	\begin{center}
		\caption{Impact of tariffs on US imports from Spain. Product-level analysis}
		\label{tab:product_US_ESP}
		{
\def\sym#1{\ifmmode^{#1}\else\(^{#1}\)\fi}
\begin{tabular}{l*{5}{c}}
\hline\hline
                    &\multicolumn{2}{c}{PPML}               &\multicolumn{3}{c}{OLS}                                    \\\cmidrule(lr){2-3}\cmidrule(lr){4-6}
                    &\multicolumn{1}{c}{(1)}&\multicolumn{1}{c}{(2)}&\multicolumn{1}{c}{(3)}&\multicolumn{1}{c}{(4)}&\multicolumn{1}{c}{(5)}\\
                    &\multicolumn{1}{c}{Value}&\multicolumn{1}{c}{Value}&\multicolumn{1}{c}{Value}&\multicolumn{1}{c}{Quantity}&\multicolumn{1}{c}{Price}\\
\hline
Tariff x Post       &      -0.781\sym{a}&      -0.762\sym{a}&      -0.529\sym{a}&      -0.509\sym{a}&      -0.020       \\
                    &     (0.266)       &     (0.257)       &     (0.124)       &     (0.128)       &     (0.046)       \\
\hline
Observations        &      339220       &      140340       &      140340       &      140340       &      140340       \\
Pseudo-R2           &       0.744       &       0.778       &           .       &           .       &           .       \\
Adj.-R2             &           .       &           .       &       0.604       &       0.707       &       0.774       \\
\hline\hline
\end{tabular}
}

		\caption*{\begin{footnotesize}Note: Regressions estimated with monthly export data from January 2017 to February 2021. The dependent variables are import value in columns~1 and~2, and log import value, log import quantity, and log import price in columns~3,~4 and~5, respectively. All regressions include product$\times$month and year fixed effects. Tariff$\times$Post turns 1 if a product was subject to a tariff and the month-year combination is within the tariff-imposition period. Standard errors clustered at the product level are in parentheses. a represents statistical significance at the 1\% level.
		\end{footnotesize}}
	\end{center}
\end{table}

I estimate Equation~\eqref{eq:product_analysis1} using a Poisson pseudo-maximum likelihood estimator \citep{santossilva2010ppml}. This model keeps the zero-import values in the estimation sample and addresses ordinary least squares (OLS) estimates' heteroscedasticity bias. Column~1 shows that the interaction coefficient is negative and statistically significant, indicating that imports of products targeted by tariffs decreased relative to non-targeted products after tariffs were introduced. Specifically, US imports from Spain of tariff-targeted products decreased by 54\% relative to the non-targeted ones (1-exp(-.781)).

Columns~4 and~5 decompose the impact of tariffs on the value of imports into its quantity and price components. Since this decomposition can only be performed with an OLS model, I re-estimate Equation~\eqref{eq:product_analysis1} with it.\footnote{As the OLS is a linear estimator, and errors have an expected value of zero, the sum of the quantity and price estimates are equal to the value estimate.} To compare point estimates, in column~2, I re-estimate the PPML model using the same sample used in the OLS estimations. The OLS point estimate (column~3) is similar to that reported in column~1, confirming that tariffs had a large negative effect on the value of imports. According to the difference-in-differences OLS coefficient, the value of imports of a product targeted by a tariff decreased relative to a non-targeted product by 41\% after the tariff was imposed (1-exp(-0.529)). This decrease is slightly lower than that estimated with the PPML model. Interestingly, almost all the reduction in the value of imports is explained by a reduction in quantity (column~4). The change in prices is statistically not different from zero (column~5). This latter result suggests that there was a complete pass-through of tariffs to import prices. This result is surprising given the large size of the US market. However, it is in line with previous studies that found complete pass-through of tariffs to import prices during the US-China trade war \citep{amiti2019impact,fajgelbaum2020return}. 

The period in which tariffs were in force coincided with the Covid-19 pandemic. The products targeted by a tariff could coincide with those that were negatively affected in demand due to social distancing measures. If that were the case, the interaction coefficient would be capturing the negative shock of the pandemic on the demand for tariff-targeted products rather than the negative effect of tariffs. To rule out this potential bias, I estimate a triple-difference specification, which compares the evolution of US imports from Spain with those from other countries that were not affected by tariffs. The triple-difference specification is defined as follows:

\begin{equation}
	\label{eq:product_analysis2}
	y_{kjtm}=exp[\beta(Tariff_{kj}*Post_{tm})+\gamma_{kj}+\gamma_{tj}+\gamma_{mj}+\gamma_{kt}+\gamma_{km}+\epsilon_{ktm}]
\end{equation}

where $y_{kjtm}$ is the value of US imports of product $k$ from country $j$ at year $t$ and month $m$. Note that the new equation includes five different fixed-effects combinations: product$\times$origin, year$\times$origin, month$\times$origin, product$\times$year, and product$\times$month. Now, the $\beta$ coefficient captures whether the difference between US imports of tariff-targeted products from Spain and other countries (first difference) changed relative to the difference between US imports of non-targeted products from Spain and other countries (second difference) after tariffs were imposed (third difference). To keep the sample computationally feasible, I select the top 30 origins of US imports in 2018, the year before tariffs were imposed, and estimate the model with OLS.\footnote{I exclude other EU members from the list of origin-of-imports countries because they could also be targeted by tariffs. China was also excluded from the sample because many of its products had been targeted by tariffs during the same period.} As shown in Table~\ref{tab:product_triple} in the Appendix, the triple-difference interaction term remains negative and significant, and has a point value similar to that reported in column~3 of Table~\ref{tab:product_US_ESP}. This result confirms that imports of tariff-targeted products did not decrease relative to the non-targeted ones because of Covid-19.

\subsection{Exporter-level analyses}

In this subsection, I analyze how tariffs affected Spanish exporters' revenue and probability of 
survival in the US. I estimate the following specification:
\begin{equation}
	\label{eq:did}
	x_{ftm}=exp[\beta(ShareTariff_{f}*Post_{tm})+\alpha_{1}Medium_{ft}+\alpha_{2}Large_{ft}+\gamma_{f}+\gamma_{tm}+\epsilon_{ftm}]
\end{equation}
where $x_{ftm}$ is the export revenue of firm $f$ in year $t$ and month $m$ in the US. $ShareTariff_{f}$ is the share of tariff-targeted products in firm $f$'s total exports to the US before tariffs were introduced. It captures the extent to which an exporter was affected by the tariffs imposed due to the Airbus-Boeing conflict. The specification also introduces two dummy variables controlling for the export-size of the firm in year $t$. Export size is defined as total exports of firm $f$ at year $t$. $Medium_{ft}$ turns 1 if the export size of the firm is above the median and below or equal to the 75 percentile in year $t$. $Large_{ft}$ turns 1 if the export size of the firm is above the percentile 75. $\gamma_{f}$ is a firm fixed effect, $\gamma_{tm}$ is a year$\times$month fixed effect, and $\epsilon_{ft}$ is the disturbance term. The key variable is the difference-in-differences interaction $ShareTariff_{f}*Post_{t}$, which captures whether the dependent variable changed for Spanish firms exporting tariff-targeted products to the US relative to the one of other Spanish exporters to the US after tariffs were introduced.  

Data on Spanish exporters were collected from the Customs and Excise Department of the Spanish Tax Agency and included the universe of Spanish exporters of goods. The data set contains a firm identifier, export destination, the product classification, the value of exports, and exported quantities. The product classification used by Spanish customs, the Combined Nomenclature, is equivalent to the US HTS classification at the 6-digit level only. I define that a 6-digit product was targeted by a tariff if any of its HTS 8-digit varieties was targeted by a tariff.\footnote{When calculating the firm-level exports, I excluded all products that were threatened by a tariff, but a tariff was not imposed. Products related to the aircraft industry were excluded as well. It is reasonable that these manufacturers anticipated a raise in tariffs on their products in the US after the WTO ruling in May 2018 and, hence, had enough time to prepare for this change in trade policy.}

\begin{figure}[htbp]
	\begin{center}
		\caption{Pre-trends. Median exports to the US by a tariff-targeted firm vs a matched firm, 2017-2021 (thousand euros; rolling 12-month sum)}
		\label{fig:pretrends_matched_median_0}
		\includegraphics[height=3.5in]{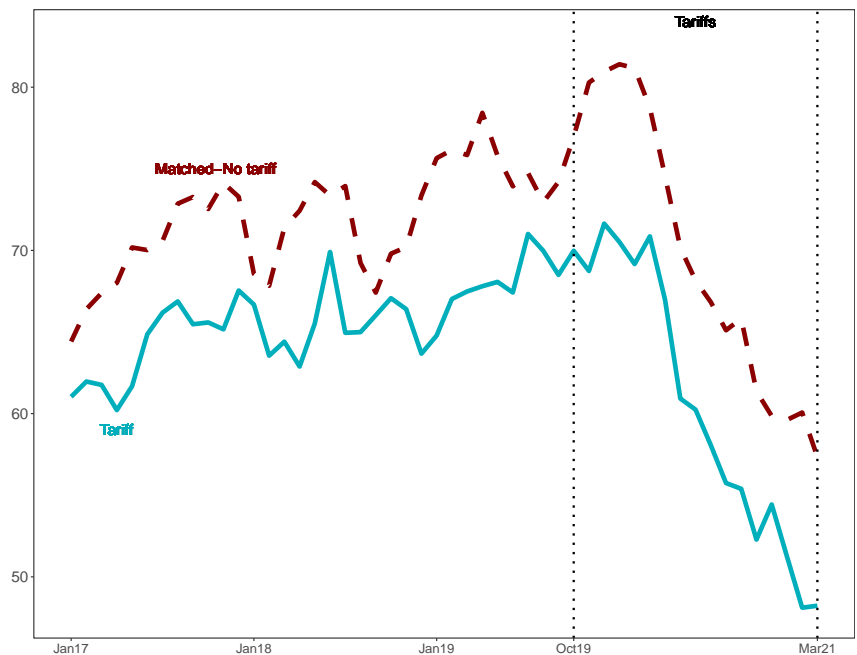}
	\end{center}
	\footnotesize Source: author's own elaboration based on Customs data.
\end{figure}

The key identifying assumption of the difference-in-differences strategy is that other exporters provide an appropriate counterfactual of the trend that exporters hit by tariffs would have followed had the US not introduced the tariffs. To ensure parallel pre-trends, I use the Mahalanobis distance algorithm to match each tariff-targeted exporter with a non-targeted exporter using export values to the US in 2016, 2017, and 2018. Figure~\ref{fig:pretrends_matched_median_0} presents the evolution of exports to the US of the median firm targeted by tariffs (solid blue line) and the median matched firm (dashed red line). By construction, the trend followed by the two types of firms is very similar in the three years before the introduction of tariffs. After tariffs were introduced, the trends are also similar: exports decrease for both tariff-targeted and non-targeted exporters.

Table~\ref{tab:summary} presents descriptive statistics of Spanish exporters of products that were targeted by tariffs and the matched exporters. They are calculated using data from 2016 to 2018, before tariffs were introduced. The number of exporters that were affected by tariffs was 1,216. Wine was the most important product for 78\% of the exporters that were hit by a tariff in the US; olive oil was the most important product for 8\% of the exporters, cheese for 3\%, and olives for 3\%. These exporters were matched with 964 firms that were not hit by tariffs. Due to the criteria used for matching, the median value of exports to the US for both groups is similar. This is also true for the number of products that both groups sell in the US market. However, the overall export revenue of the tariff-targeted exporters is smaller than that of matched exporters. The matched firms also export to more destinations and more products than the tariff-targeted firms. Interestingly, the importance of the US market in total exports for the firms that were hit by tariffs was larger than for the matches.
\begin{table}[tbp] \centering
\newcolumntype{C}{>{\centering\arraybackslash}X}

\caption{Tariff-targeted exporters vs. matched exporters to the US before tariffs}
\label{tab:summary}
{\small
\begin{tabularx}{\textwidth}{lrr}

\toprule
\multicolumn{1}{l}{}&\multicolumn{1}{p{2.5cm}}{\raggedleft Tariff-targeted exporters}&\multicolumn{1}{p{2.6cm}}{\raggedleft Matched exporters} \tabularnewline
\midrule\addlinespace[1.5ex]
Number of exporters to the US&1216&964 \tabularnewline
Export revenue (thousand euros)&233&1458 \tabularnewline
Export revenue in the US (thousand euros)&54&57 \tabularnewline
Number of export destinations&6&13 \tabularnewline
Number of exported products&3&6 \tabularnewline
Number of exported products to the US&2&2 \tabularnewline
Exports to the US as a share of total exports (\%)&33&8 \tabularnewline
\bottomrule \addlinespace[1.5ex]
\end{tabularx}
}
\footnotesize \raggedright Note: average yearly values for the period 2016-2018. A tariff-targeted exporter is a firm that exports at least one product that was targeted by tariffs. Matched exporters are firms that exported no products hit by tariffs.

\end{table}

Table~\ref{tab:int_ext_us} presents the estimations of Equation~\eqref{eq:did} when export revenue is the dependent variable. I estimate a Poisson pseudo-maximum likelihood model. Column~1 shows that the interaction coefficient, although negative, is statistically insignificant. In line with the visual impression from Figure~\ref{fig:pretrends_matched_median_0}, this result indicates that the export revenue of firms that exported products targeted by tariffs did not significantly decrease relative to the one of the control firms after tariffs were introduced. 

\setlength{\tabcolsep}{3.5pt}
\begin{table}[htbp]
	\begin{center}
		\footnotesize
		\caption{Impact of tariffs on Spanish exporters to the US}
		\label{tab:int_ext_us}
		{
\def\sym#1{\ifmmode^{#1}\else\(^{#1}\)\fi}
\begin{tabular}{l*{4}{c}}
\hline\hline
                    &\multicolumn{3}{c}{PPML}                                   &\multicolumn{1}{c}{OLS}\\\cmidrule(lr){2-4}\cmidrule(lr){5-5}
                    &\multicolumn{1}{c}{(1)}&\multicolumn{1}{c}{(2)}&\multicolumn{1}{c}{(3)}&\multicolumn{1}{c}{(4)}\\
                    &\multicolumn{1}{c}{Value}&\multicolumn{1}{c}{Value}&\multicolumn{1}{c}{Value}&\multicolumn{1}{c}{Exit}\\
\hline
Share tariff x Tariff period&      -0.124       &      -0.105       &      -0.119       &       0.015       \\
                    &     (0.111)       &     (0.100)       &     (0.115)       &     (0.016)       \\
[1em]
Tariff period 3 lags&                   &       0.040       &                   &                   \\
                    &                   &     (0.097)       &                   &                   \\
[1em]
Tariff period 2 lags&                   &      -0.025       &                   &                   \\
                    &                   &     (0.065)       &                   &                   \\
[1em]
Share tariff x Potential period&                   &                   &       0.045       &                   \\
                    &                   &                   &     (0.075)       &                   \\
\hline
Observations        &      114106       &       97168       &      114106       &        8587       \\
Pseudo-R2           &       0.858       &       0.862       &       0.858       &           .       \\
Adj.-R2             &           .       &           .       &           .       &       0.105       \\
\hline\hline
\end{tabular}
}

		\caption*{\begin{footnotesize}Note: The dependent variable is monthly exports to the US in columns~1,~2, and~3. In column~4 the dependent variable turns if the firm ceases to export to the US. All regressions include firm and year$\times$month fixed effects, and year-specific dummies to control for firm size. Standard errors clustered at the firm level are in parentheses.
	\end{footnotesize}}
	\end{center}
\end{table}
\setlength{\tabcolsep}{6pt}


As explained above, the majority of exporters that were hit by the increase in tariffs were wine exporters. Hence, the insignificant effect of tariffs on export revenue reported in column~1 may be reflecting the particular situation in that industry. To determine whether tariffs had a similar effect in other industries, I classify exporters according to their most important product, and estimate Equation~\eqref{eq:did} for each group. Table~\ref{tab:int_categories}, in the Appendix, reports the results for the top-4 Spanish product categories targeted by a tariff: olive oil, wine, olives, and cheese. In all products, the export revenue in the US of the exporters targeted by tariffs did not decrease significantly relative to their matched exporters after tariffs were imposed. 

To further test the validity of the parallel pre-trends' assumption, I divide the period before the imposition of tariffs into three sub-periods. The first sub-period, Tariff period 1 lag, covers the October 2018-September 2019 period; Tariff period 2 lags covers the October 2017-September 2018 period, and the Tariff period 3 lags covers the October 2016-September 2017 period. Column~2 presents the results of estimating Equation~\eqref{eq:did} with Tariff period 3 lags and Tariff period 2 lags as additional variables. Tariff period 1 lag is the reference category. The Tariff period 3 lags and Tariff period 2 lags coefficients are statistically insignificant, confirming that exporters that were hit by tariffs had a similar trend to the one of exporters that were not hit by tariffs before they were applied.

In column~3, I analyze whether the export revenue changed between the date in which the potential list of targeted products was announced (April 2019) and the final list of targeted products was published (October 2019). The Share tariff x Potential period interaction is statistically insignificant, indicating that the export revenue of potentially-targeted exporters did not change relative to the one of other exporters to the US during the period in which the potential products that could be targeted by a tariff was known.


Next, I analyze whether the imposition of tariffs increased the risk of ceasing to export to the US for firms that were hit by tariffs. I estimate the following equation:
\begin{equation}
	\label{eq:did_exit}
	Exit_{fz}=\beta(ShareTariff_{f}*Post_{z})+\alpha_{1} Medium_{fz}+\alpha_{2}Large_{fz}+\gamma_{f}+\gamma_{z}+\epsilon_{fz}
\end{equation}
where $Exit_{fz}$ turns 1 if the firm exited the US market during period $z$. To estimate Equation~\eqref{eq:did_exit}, I divide the period of analysis into four sub-periods. Since tariffs were in force for 16 full months (November 2019-February 2021), I define three 16-months sub-periods for the pre-tariff period: 1) October 2015-January 2017; 2) February 2017-May 2018, and 3) June 2018-September 2019. For each sub-period, I compute whether the firm exported to the US and estimate a linear probability model. Column~4 of Table~\ref{tab:int_ext_us} shows that the risk of ceasing to export to the US did not increase significantly for exporters that were hit by tariffs relative to other exporters after tariffs were introduced.

I also analyzed whether exporters hit by tariffs in the US changed their export performance relative to other Spanish exporters to the US in third foreign markets. I do not find any significant change between the two groups of exporters.\footnote{To conserve space, estimates are available upon request.}

In sum, the firm-level analysis indicates that Spanish exporters targeted by tariffs had similar economic outcomes in the US as not-targeted exporters. The next section analyzes how Spanish exporters targeted by tariffs were able to neutralize the negative impact of the tariff hike.

\section{Spanish exporters' tariff-avoidance strategies}
\label{sec:tariff_avoidance_strategies}

Spanish exporters followed two major strategies to counteract the introduction of tariffs: 1) the substitution of Spanish products by products originated in countries that had not been affected by a tariff; and 2) the substitution of tariff-targeted varieties by non-targeted ones. This section describes these tariff-avoidance strategies.

\subsection{Substitution of Spanish products by products originated in non-targeted countries}
The first tariff-avoidance strategy followed by Spanish exporters to the US was to substitute a Spanish product targeted by a tariff with the same product originated in a non-targeted country. This strategy was used by olive oil exporters. 

\begin{figure}[htbp]
	\begin{center}
		\caption{Imports of olive oil by Spanish exporters of olive oil to the US, 2019-2022 (Million euros. Rolling 12-month sum)}
		\label{fig:m_oliveoil_Spanish_exporters}
		\includegraphics[height=3.5in]{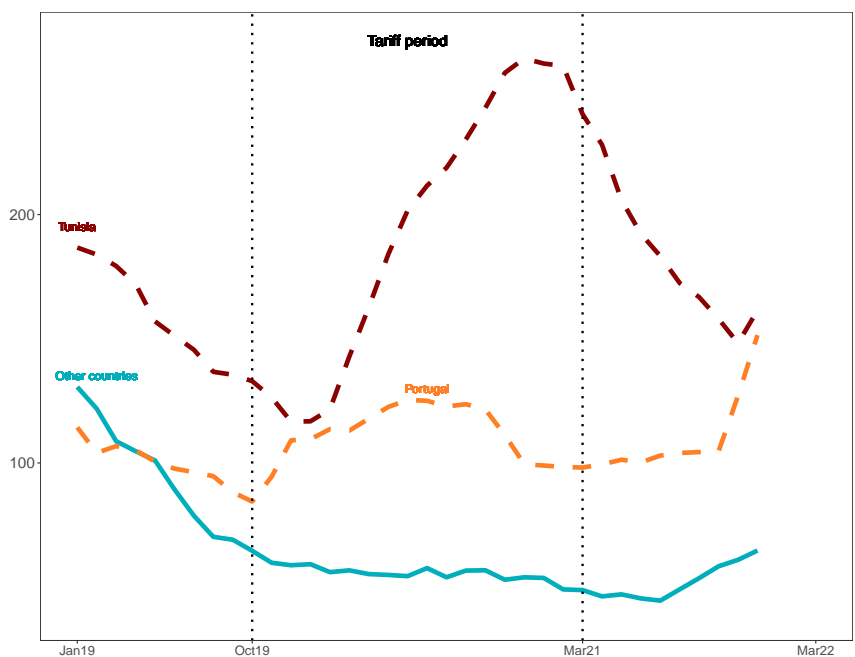}
	\end{center}
	\footnotesize Source: author's own elaboration based on US Customs data.
\end{figure}

Figure~\ref{fig:m_oliveoil_Spanish_exporters} shows the imports of olive oil by the Spanish olive oil exporters to the US. These firms accounted for 86\% of all Spanish olive oil imports in 2018, the year before tariffs were introduced. Tunisia was the most important origin of imports for these firms in 2018, accounting for 41\% of total imports, followed by Portugal with 27\%. These two countries were not targeted by a tariff on olive oil. There was a decrease in the value of imports from all countries until tariffs were imposed by the US. During the tariff period, Spanish exporters of olive oil to the US increased imports from Portugal and, specially, from Tunisia. There is also a very sharp decrease in the value of imports from Tunisia once tariffs were lifted.

As reported by the media and confirmed by the author in personal communications with industry representatives, a large share of the olive oil in bulk imported from Tunisia and Portugal during the tariff period was bottled in Spain and exported to the US in substitution of the tariff-targeted olive oil originated in Spain.\footnote{In the media, see, among others, \url{https://www.elconfidencial.com/economia/2020-10-10/espana-aceite-trump-compra-otros-paises_2782899/} and \url{https://www.elconfidencial.com/espana/andalucia/2021-02-17/aceite-envasado-espana-vendio-eeuu-tunez-portugal_2954147/}. I interviewed Rafael Pico, Director of Asoliva, Spanish Association of Industry and Exporters of Olive Oil.} The label of these olive oil containers exported by Spanish firms to the US stated that the origin of the product was Tunisia or Portugal. This was key for Spanish exporters' tariff-avoidance strategy. US customs considered that the bottling that had taken place in Spain did not constitute a major product transformation. Hence, US Customs recorded the shipments from Spain as imports of olive oil from Tunisia or Portugal. Since these countries had not been targeted by a tariff, Spanish exporters could avoid the additional import duties.

\begin{figure}[htbp]
	\begin{center}
		\caption{US imports of olive oil by countries, 2019-2022 (Million USD. Rolling 12-month sum)}
		\label{fig:m_oliveoil_usa}
		\includegraphics[height=3.5in]{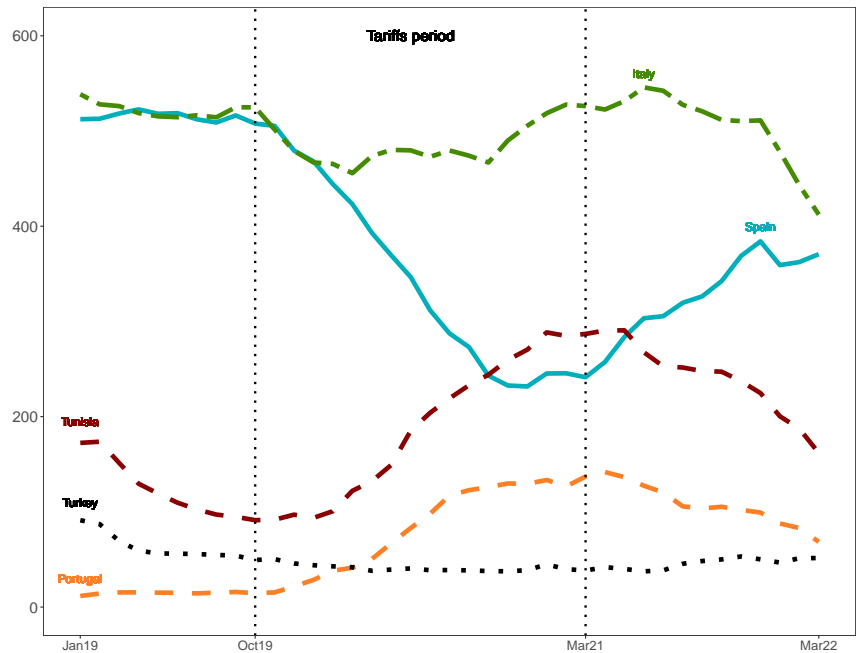}
	\end{center}
	\footnotesize Source: author's own elaboration based on US Customs data.
\end{figure}

Figure~\ref{fig:m_oliveoil_usa} shows the evolution of US imports of olive oil between 2019 and the first quarter of 2022 from its top-4 suppliers in 2018: Italy, Spain, Tunisia, and Turkey, which accounted for 89\% of total olive oil imports in that year. It also plots imports from Portugal, the top-9 supplier, which accounted for 1\% of total imports in 2018. There is an increase in olive imports from Portugal and, particularly, from Tunisia after tariffs were imposed. Imports from these countries decreased once tariffs were lifted. There is no significant change in the value of imports from Italy and Turkey, countries that were not targeted by tariffs and occupied the top and the top-4 position, respectively as suppliers of olive oil to the US in 2018. These trends are in line with the narrative that Spanish exporters substituted Spanish olive oil with olive oil originated in Tunisia and Portugal in their US exports during the period in which tariffs were in force.

\begin{figure}[htbp]
	\begin{center}
		\caption{Olive oil. US imports from Spain vs. Spanish exports to the US, 2019-2022 (rolling 12-month sum; September 2019=100)}
		\label{fig:olive_oil_m_vs_x}
		\includegraphics[height=3.5in]{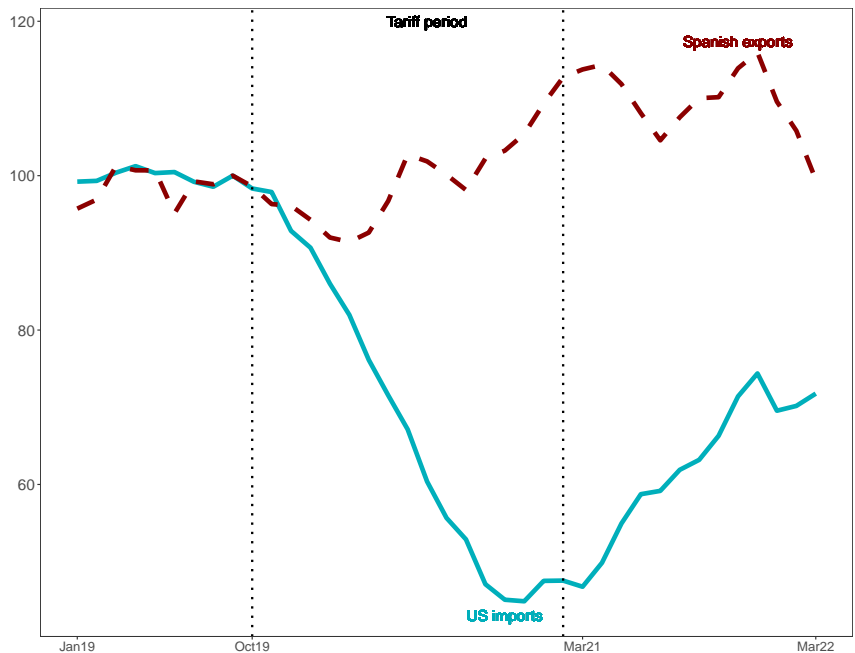}
	\end{center}
	\footnotesize Source: author's own elaboration based on Customs and US Census data.
\end{figure}

The tariff-avoidance strategy of Spanish olive oil exporters also led to a large discrepancy between the value of exports reported by Spanish customs and the mirror value of imports reported by US customs. The dashed red line in Figure~\ref{fig:olive_oil_m_vs_x} captures the rolling 12-month sum value of exports of olive oil from Spain to the US reported by Spanish customs. The solid blue line is the mirror imports reported by US Customs. I set the value of these flows at 100 in September 2019, the month before tariffs were introduced.\footnote{To make the figures comparable, the Spanish tariff-targeted olive oil exports and the mirror US imports are calculated at the HS 6-digit level.}  The Spanish and US lines are close and move parallel in the months before the introduction of tariffs. However, since tariffs were introduced the two lines move in opposite directions: US imports decrease whereas Spanish exports rise. The gap between the lines is remarkable. By the end of the tariff period, US customs had recorded a 53\% decline in the value of olive oil imports from Spain relative to the pre-tariff period, whereas Spanish customs had recorded a 14\% increase in the value of olive oil exports to the US.

This difference is explained by the fact that Spanish Customs recorded as exports to the US all containers shipped from Spain to the US, although a large share of them had not been filled with olive oil produced in Spain. This recording criteria explains why Spanish exports of olive oil to the US did not decline, and even increased, despite the fact that an additional 25\% tariff had been imposed. It also provides the first explanation on why a large decline is observed in US imports of Spanish tariff-targeted products (Figure~\ref{fig:esp_tar_vs_rest}) and, at the same time, we find no significant effect on the export revenue of Spanish exporters of tariff-targeted products to the US relative to other Spanish exporters (Table~\ref{tab:int_ext_us}). Note that the gap between Spanish exports and the mirror US imports began to narrow once tariffs were lifted in March 2021. However, a large gap remained even one year after tariffs had been removed. This suggests that some Spanish exporters continued to export olive oil originated in other countries after the elimination of tariffs.

As mentioned in the introductory section of the paper, many studies have relied on the gap between the value of exports and the mirror value of imports to identify tariff-evasion. The present case highlights that the gap can also be the outcome of exporters' tariff-avoidance strategies and the differences in the criteria to record trade flows between the exporting and importing customs offices.

As additional evidence, I ran a simple OLS regression:
\begin{equation}
\label{eq:simple_ols}
x_{i}=\alpha+\beta\Delta m_{i}+\epsilon_{i}
\end{equation}

where $x_{i}$ is the value of olive oil exports to the US by firm $i$ during the tariff period, $\alpha$ is a constant, $\Delta m_{i}$ is the change in imports of olive oil from Tunisia and Portugal between the tariff and pre-tariff period, and $\epsilon_{i}$ is the disturbance term. 

The $\beta$ coefficient is expected to be positive: only Spanish olive oil exporters that increased imports from Tunisia and Portugal could substitute the Spanish olive oil with the olive oil of non-targeted countries and keep their export revenue in the US. Table~\ref{tab:simple_ols} in the Appendix reports a positive and statistically significant $\beta$ coefficient: a 1000 euro increase in olive oil imports from Tunisia and Portugal was correlated with a 627 euros increase in olive exports to the US market. This result is in line with the argument that Spanish olive oil exporters substituted the Spanish olive oil with olive oil from non-targeted countries to avoid the negative impact of tariffs on their exports to the US.


The substitution of olive oil produced in Spain with olive oil originated in non-tariff countries was a feasible strategy for Spanish exporters whose competitive strategy in the US was based on prices. The main goal of these exporters was to maintain their relationship with large US retailers. These retailers would not have accepted a rise in prices, forcing the Spanish exporters to find alternatives to offer a price-competitive product.\footnote{For example, the export manager of a major Spanish olive oil firm asserted that large US distributors would not have allowed even a 2\% increase in prices (\url{https://issuu.com/olimerca/docs/vermultimedia/s/10169409}). In another related context, \cite{iacovone2015walmart} showed that Walmart demanded a lower price than traditional retailers to its suppliers in exchange of a larger market in Mexico. In contrast, results by \cite{cavallo2021passthrough} suggested that in the US-China trade war retailers absorbed a large share of the increase in costs due to tariffs. They explained this fact based on the differentiated nature of the products that the US imported from China.} However, product substitution was not an option for exporters that had built their competitive strategy on the relationship between a high-quality olive oil and its Spanish origin.

Spanish firms could also use two additional strategies to avoid tariffs in the US. Firstly, olive oil was not targeted by tariffs if it was exported in bulk. This benefited Spanish exporters that were already bottling olive oil in the US. Bottling in the US might become an option for the rest of exporters. However, US imports of olive oil in bulk from Spain decreased during the tariff period, indicating that Spanish exporters did not follow this strategy. The main reason was that the cost of bottling olive oil in the US was too high for many exporters.\footnote{\url{https://issuu.com/olimerca/docs/vermultimedia/s/10169409}.} Secondly, Spanish exporters that already had production facilities in countries that had not been targeted by tariffs (e.g., Portugal and Tunisia) increased their exports to the US directly from those countries.\footnote{\url{https://www.elconfidencial.com/empresas/2019-10-04/aceite-oliva-arancel-aceituna-exportacion_2268007/}.}

\begin{figure}[htbp]
	\begin{center}
		\caption{Analysis by product: US imports from Spain vs. Spanish exports to the, 2019-2021 (rolling 12-month sum; September 2019=100)}
		\label{fig:esp_vs_usa_hs6_product}
		\begin{tabular}{cc}
			{\fontsize{10}{11}\selectfont Olive oil}&{\fontsize{10}{11}\selectfont Wine}\\
			\includegraphics[scale=0.275]{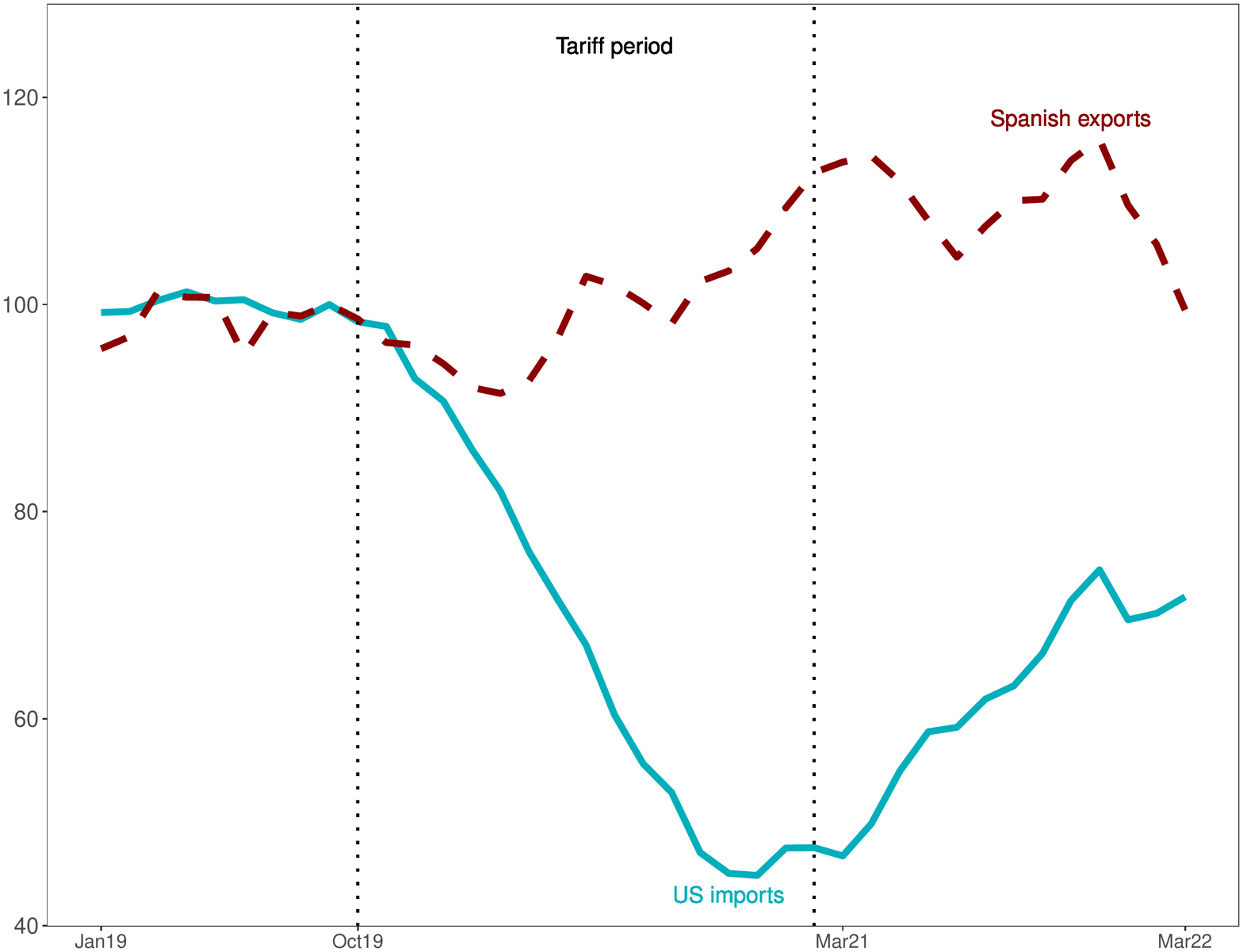}&		\includegraphics[scale=0.275]{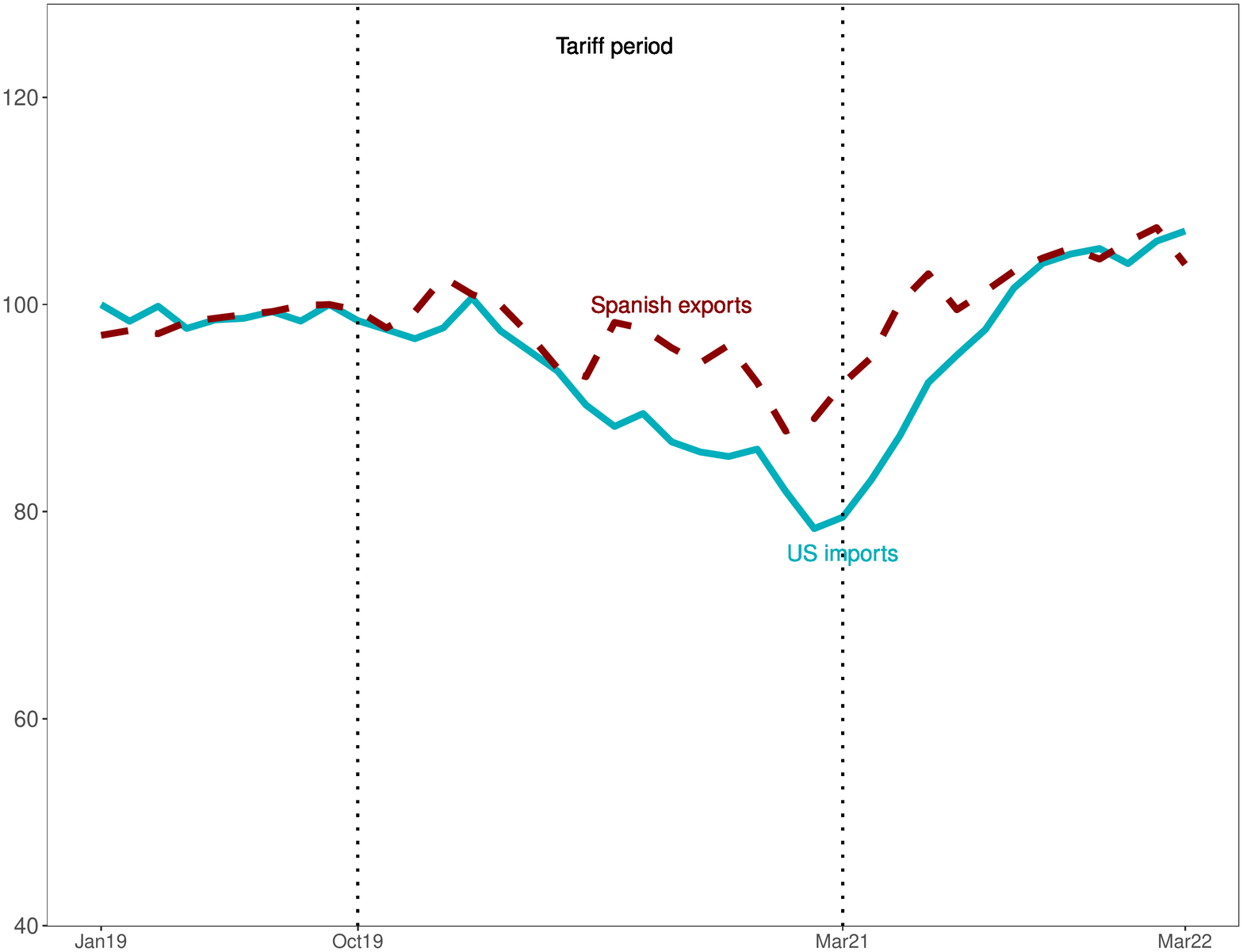}\\
			[1em]
			{\fontsize{10}{11}\selectfont Olives}&{\fontsize{10}{11}\selectfont Cheese}\\
			\includegraphics[scale=0.275]{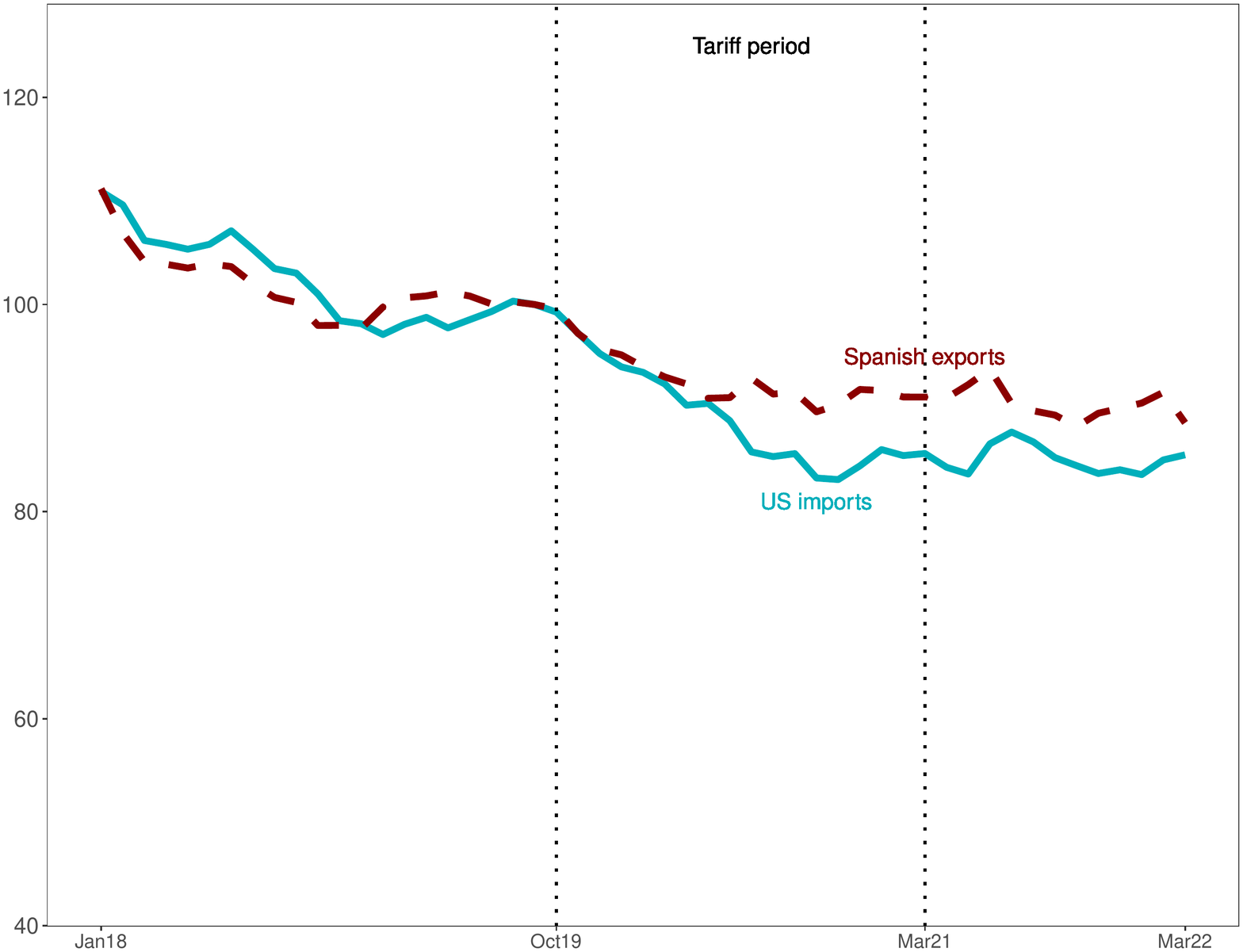}&		\includegraphics[scale=0.275]{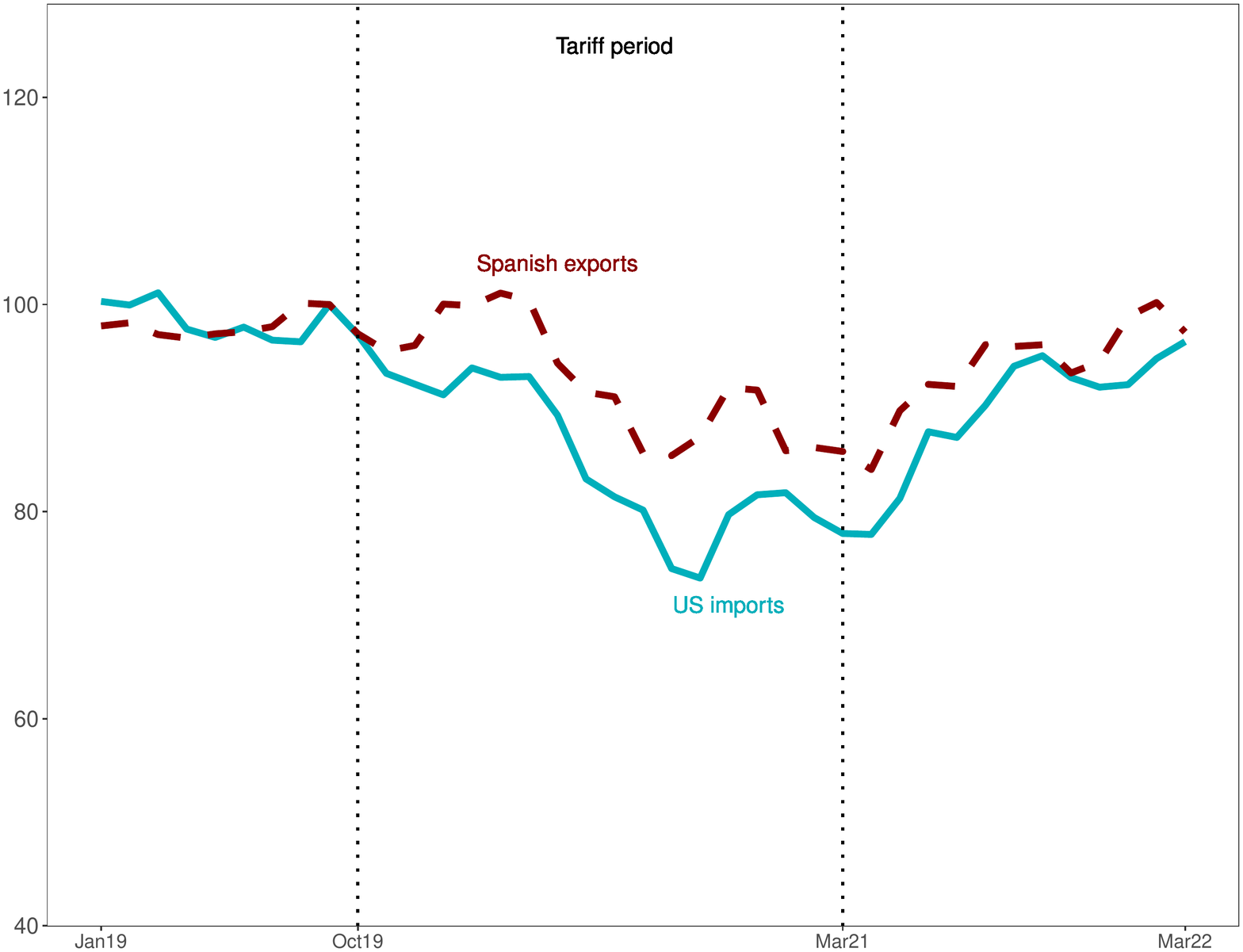}\\
			
		\end{tabular}
	\end{center}
	\footnotesize Source: author's own elaboration based on US Census and Agencia Tributaria trade data.
\end{figure}

Figure~\ref{fig:esp_vs_usa_hs6_product} shows that the gap between the value of Spanish exports to the US and the mirror value of US imports from Spain also widened during the tariff period for wine, olives, and cheese. However, the increase in those products was negligible compared to that in the olive oil industry. One explanation for these differences lies in product regulations. For example, most of the wine imported by the US from Spain are of varieties protected with designation of origin. Hence, these exporters cannot substitute the origin of their product. Another explanation, as analyzed in the next subsection, is that, in the case of wine and cheese, Spanish exporters had an alternative strategy to avoid tariffs.

\subsection{Substitution of tariff-targeted varieties with non-targeted ones}

\begin{figure}[htbp]
	\begin{center}
		\caption{Substitution of tariff-targeted varieties with non-targeted ones in US imports from Spain, 2019-2021 (rolling 12-month sum; million USD)}
		\label{fig:tariff_vs_nontariff_hs6}
		\begin{tabular}{cc}
			{\fontsize{10}{11}\selectfont Olive oil}&{\fontsize{10}{11}\selectfont Wine}\\
			\includegraphics[scale=0.275]{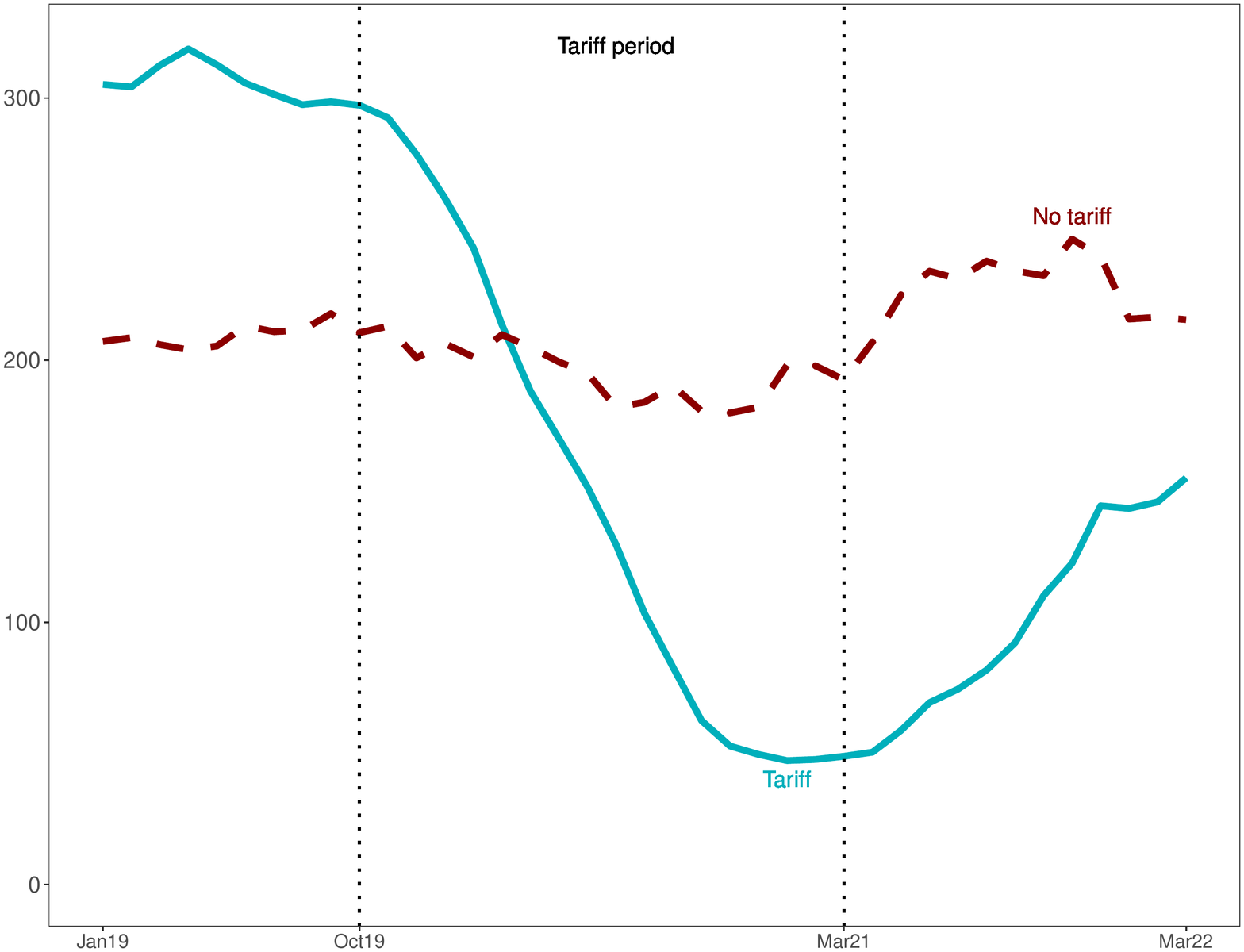}&		\includegraphics[scale=0.275]{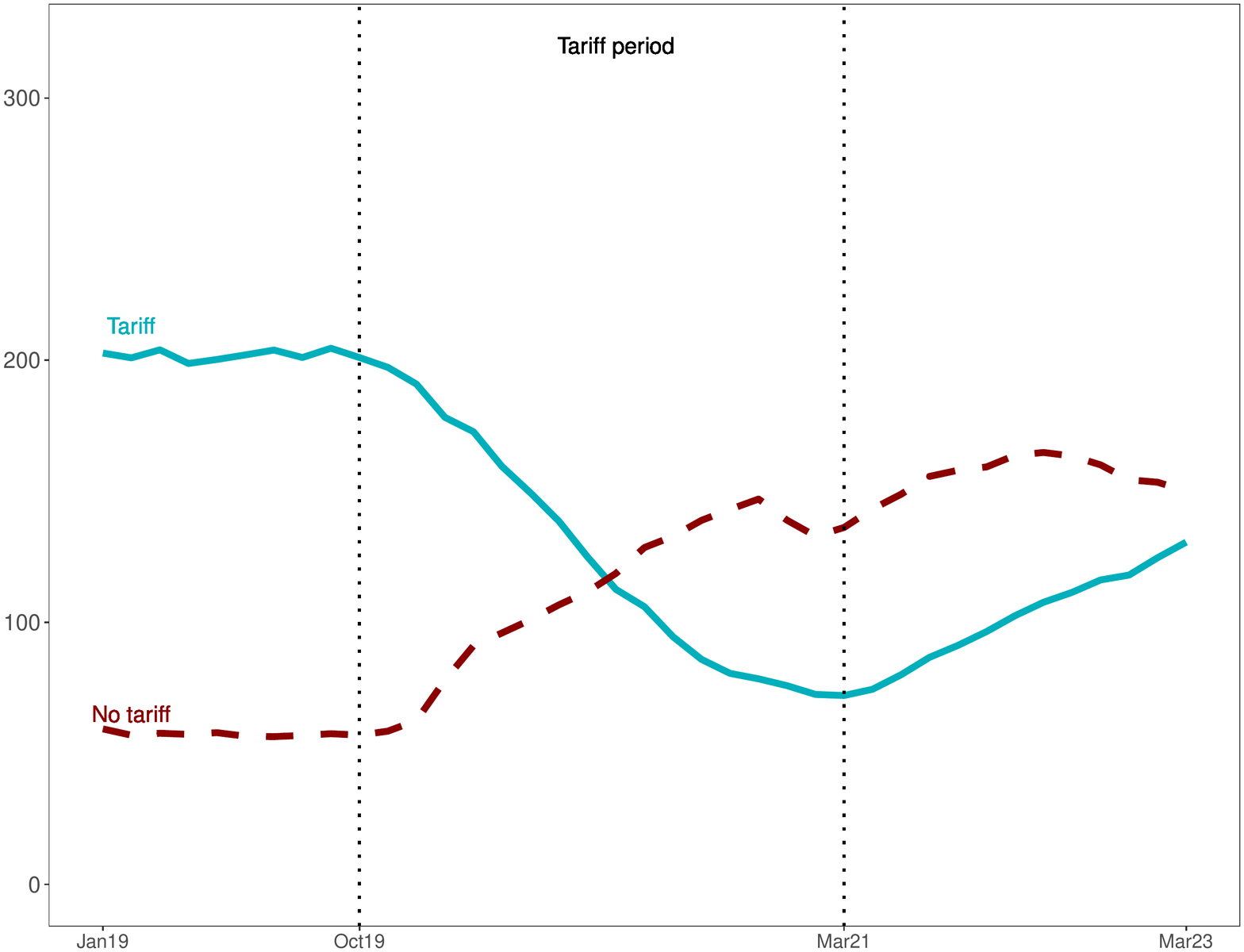}\\
			[1em]
			{\fontsize{10}{11}\selectfont Olives}&{\fontsize{10}{11}\selectfont Cheese}\\
			\includegraphics[scale=0.275]{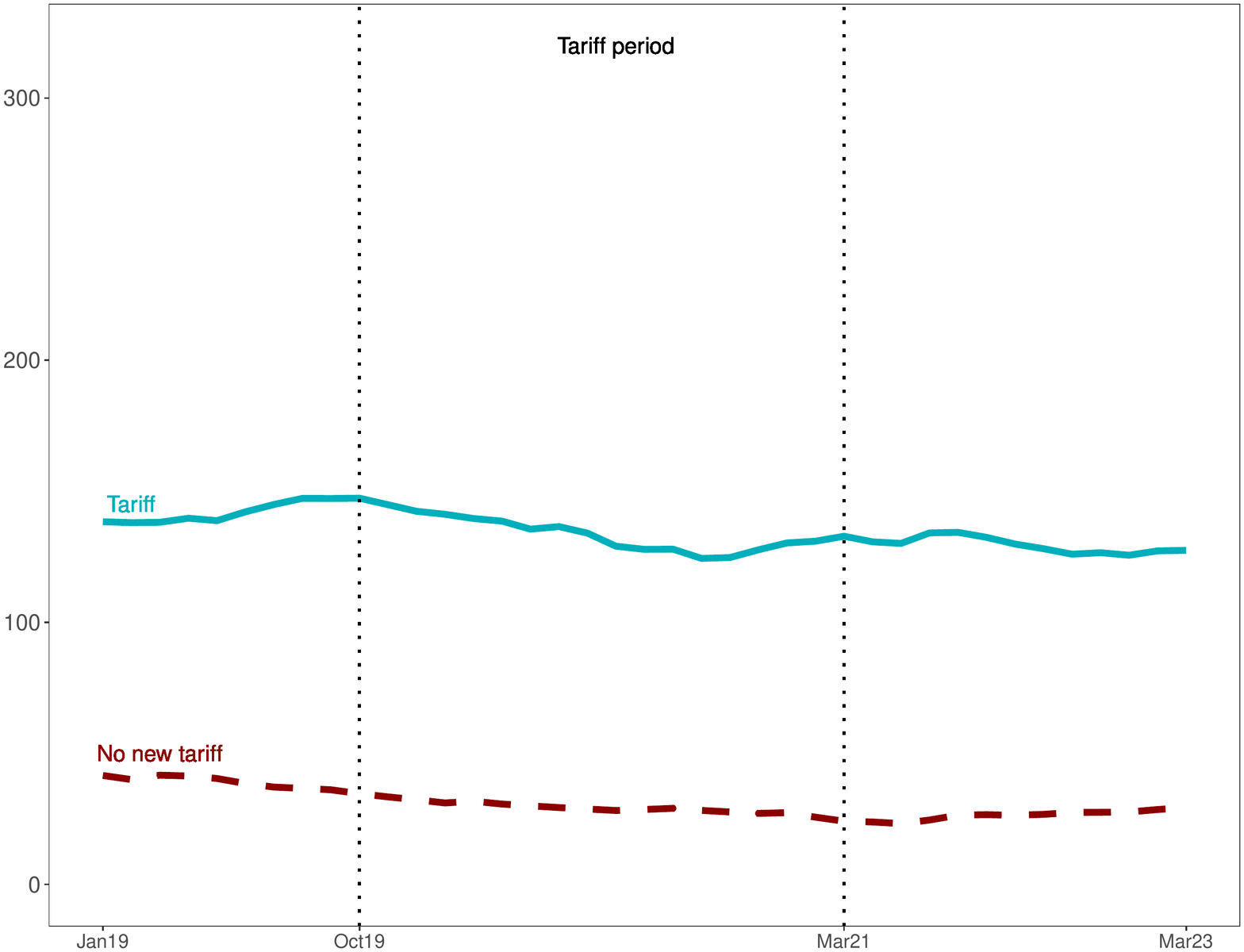}&		\includegraphics[scale=0.275]{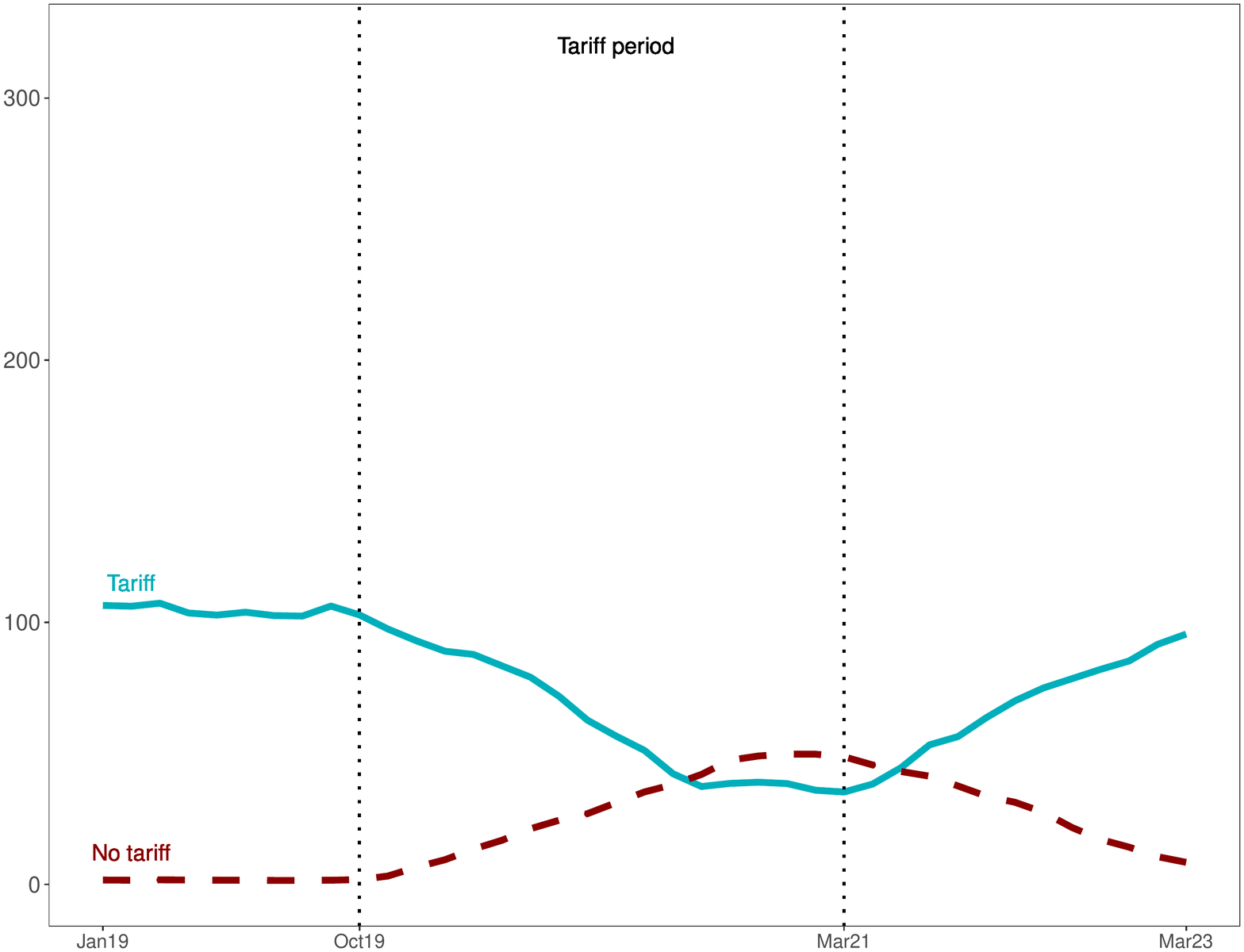}\\
		\end{tabular}
	\end{center}
	\footnotesize Source: author's own elaboration based on US Census trade data.
\end{figure}

The second strategy followed by Spanish exporters to avoid the negative effects of tariffs was to substitute the tariff-targeted varieties with similar non-targeted ones. Figure~\ref{fig:tariff_vs_nontariff_hs6} compares the evolution of US imports from Spain of tariff-targeted varieties and non-targeted varieties in the top-4 products targeted by tariffs. For each HTS 8-digit variety targeted by a tariff, I identify other varieties belonging to the same HTS 6-digit product that were not targeted by a tariff. There is a decrease in US imports of the tariff-targeted varieties when tariffs enter into force in the four products analyzed. For wine and cheese, there is a parallel increase in imports of non-targeted varieties. In the case of wine, the increase in imports of non-targeted varieties covered 57\% of the decrease in imports of targeted varieties (75 out of 132 million USD); in the case of cheese, 80\% of the decrease was covered by non-targeted varieties (48 out of 60 million USD). 

Contrary to wine and cheese, there is no substitution in imports between targeted and non-targeted varieties in olive oil and olives. In the former, as explained above, firms could have avoided the tariff exporting olive oil in bulk. However, firms did not use this strategy, because it was costly and they had the alternative of exporting olive oil originated in non-targeted countries. In the latter, it was the green-olive variety that was targeted by the additional tariff. The other main variety, black olive, had already been targeted by countervailing and anti-dumping measures since August 2018.\footnote{The combined tariff equivalent of these measures was similar to the 25\% additional tariff applied on the green variety. See \url{https://www.govinfo.gov/content/pkg/FR-2018-08-01/pdf/2018-16449.pdf} and \url{https://www.govinfo.gov/content/pkg/FR-2018-08-01/pdf/2018-16450.pdf}.} Hence, shifting from targeted to non-targeted varieties was not an option for olive exporters. It is interesting to observe that the relative decrease in imports after tariffs were introduced was smaller in olives than in the rest of products. I address this fact at the end of this section.

\begin{figure}[htbp]
	\begin{center}
		\caption{Wine. US imports by alcoholic strength, 2019-2022 (rolling 12-month sum; million USD)}
		\label{fig:substitution_wine}
	\includegraphics[height=3.5in]{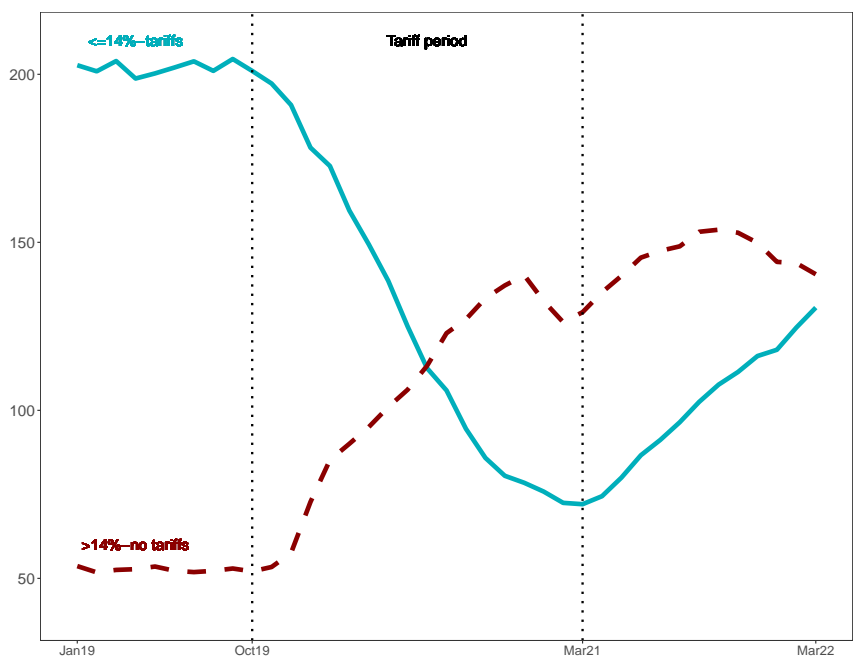}
	\end{center}
	\footnotesize Source: author's own elaboration based on US Census trade data.
\end{figure}

The substitution of the exported varieties in wine and cheese was feasible because some targeted and non-targeted varieties were similar. Regarding wine, the variety targeted by the tariff was wine in containers holding 2 liters or less with an alcoholic strength by volume not over 14\%. However, the variety with an alcohol strength over 14\% in the same type of containers was not targeted by tariffs. Figure~\ref{fig:substitution_wine} shows that there was substitution in imports between these varieties once tariffs were introduced. 

Spanish red wine, which accounted for 75\% of total Spanish exports of wine in containers holding 2 liters or less before tariffs were introduced, has an alcohol strength in the 13\%-15\% range.\footnote{\url{https://catatu.es/blog/graduacion-vino/}.} Hence, it was feasible for Spanish red wine exporters to substitute red-wine varieties with an alcohol strength equal or below 14\% with those of an alcohol strength above 14\%. This may explain why exports of red wine in containers holding 2 liters or less only declined by 7.6\% during the tariff period. However, Spanish white wine, which accounted for 18\% of total Spanish exports of wine in containers holding 2 liters or less before tariffs were introduced, has an average alcohol strength between 10\% and 11\%. In this case, the shift to varieties with an alcohol strength above 14\% was not feasible. For this type of wine, exports declined by 23.4\% relative to the pre-tariff period.

The US also imposed an additional tariff on wines originating from France and Germany. However, contrary to Spain, the tariff was imposed on wines with an alcoholic strength equal to 14\% or less and wines with an alcoholic strength over 14\%. No substitution between wine varieties happened in US imports from these countries once tariffs were imposed.

It would have been ideal to confirm the shift between wine varieties using exporter-level data. However, this was not possible since the alcohol strength threshold used by the EU trade classification is set at 15\% rather than 14\%, concealing the shift between varieties that happens at the latter threshold. Finally, it is interesting to observe that US imports of wines with an alcoholic strength equal to or below 14\% began to increase after tariffs were lifted. Imports of varieties with an alcohol strength above 14\% only began to decline several months after tariffs were lifted.

\begin{figure}[htbp]
	\begin{center}
		\caption{Cheese. US imports from Spain by top categories, 2019-2022 (rolling 12-month sum; million USD)}
		\label{fig:substitution_cheese}
		\includegraphics[height=3.5in]{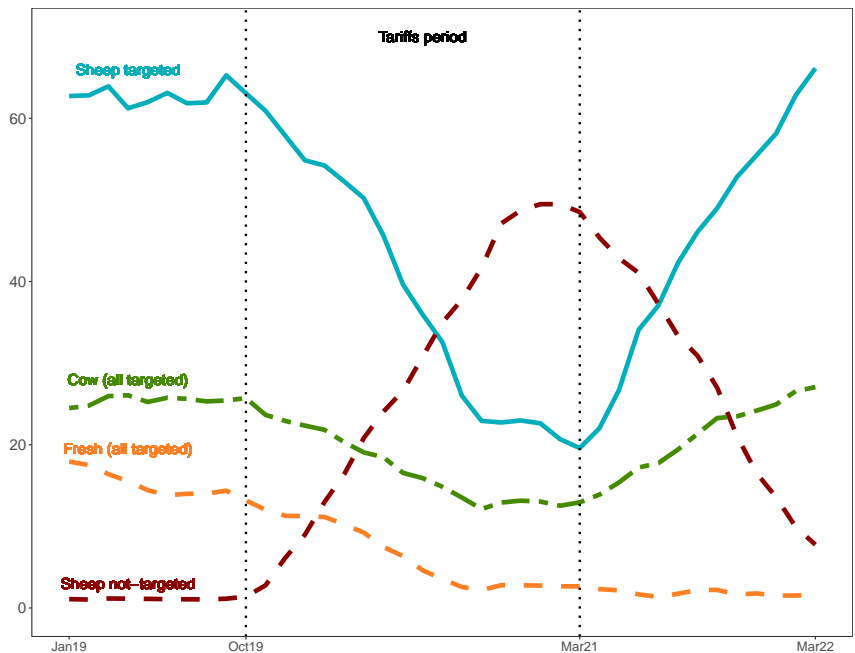}
	\end{center}
	\footnotesize Source: author's own elaboration based on US Census trade data.
\end{figure}

Figure~\ref{fig:substitution_cheese} plots the evolution of US imports of cheese from Spain before, during, and after the tariff period. Three categories account for most of imports: cheese made from sheep's milk, fresh cheese, and other cheeses containing cow's milk. Tariffs were imposed on the three varieties. However, the cheese made from sheep's milk category had a close variety which was not targeted by a tariff.\footnote{Specifically, HTS codes 04069056 and 04069057 were targeted by an additional 25\% tariff, whereas code 04069059 was not.} In the rest of the categories, all alternative varieties were targeted by a tariff. The figure shows a large decrease in imports of the cheese made from sheep's milk targeted by tariffs and a big increase in the cheese made from sheep's milk not targeted by tariffs. In fact, the non-targeted variety more than compensated for the decrease in imports in the targeted varieties. These trends suggest that Spanish firms could switch rapidly from the targeted varieties to the non-targeted ones in their exports to the US. In contrast, in the categories without an alternative non-targeted variety, the decrease in imports is less intense than in the targeted sheep category.

Note that the cheese made from sheep's milk varieties reversed their trends once tariffs were lifted. This further suggests that Spanish firms had shifted the targeted with the non-targeted varieties in their exports to avoid the negative effect of tariffs. 

As was the case for wine, it would have been ideal to explore the shift across varieties using data at the exporter level. However, the product classification used by the US at the 8-digit level, the one at which tariffs were imposed, is different to the one used by the EU at the 8-digit level disaggregation. This difference in classification precluded me from identifying the varieties targeted by tariffs and the non-targeted ones in the EU classification at the 8-digit level.

Figure~\ref{fig:m_usa_4products} and~\ref{fig:tariff_vs_nontariff_hs6} showed that the negative impact of tariffs on US imports of Spanish olives was much smaller than in other tariff-targeted products. To understand this differential impact, I estimate the following regression:  

\begin{equation}
	\label{eq:olives}
		m_{jktm}=exp[\beta(Tariff_{jk}*Post_{tm})+\gamma_{jkm}+\gamma_{t}+\epsilon_{ktm}]
	\end{equation}

where $m_{jktm}$ are US imports of olive variety $k$ from country $j$ at year $t$ and month $m$. The difference-in-differences interaction term, $Tariff_{jk}*Post_{tm}$ turns 1 if the US imposed a tariff on country $j$ and the period analyzed is when tariffs were in force. The sample covers the period January 1997-February 2021 and includes all countries the US imported olives from.\footnote{In addition to Spain, olives imported from France, Germany, and the UK were targeted by tariffs. However, the share of these countries in US total imports of tariff-targeted olives in 2018 was only 0.5\%.} The sample only includes the olive varieties that were targeted by tariffs.

Column~1 of Table~\ref{tab:reg_olives} shows that olives imports from countries targeted by tariffs decreased in the period in which the tariff was in force. However, the coefficient is statistically insignificant. This result was unexpected, since one would anticipate a significant decrease in imports from countries targeted by an additional 25\% tariff. To understand this result, it is important to consider the share of Spain in the US total olives imports targeted by tariffs. In 2018, this share was 62\%. The next countries in the ranking were Greece and Italy, with 26\% and 7\% of total US imports, respectively. Spain's large share suggests that Spanish olives were competitive in the US, difficult to substitute, and commanded a low import demand elasticity. This low demand elasticity could explain why a 25\% increase in tariffs did not lead to a significant decrease in imports.

\begin{table}[t]
	\begin{center}
		\caption{Impact of tariffs on US imports of olives}
		\label{tab:reg_olives}
		{
\def\sym#1{\ifmmode^{#1}\else\(^{#1}\)\fi}
\begin{tabular}{l*{5}{c}}
\hline\hline
                    &\multicolumn{2}{c}{PPML}               &\multicolumn{3}{c}{OLS}                                    \\\cmidrule(lr){2-3}\cmidrule(lr){4-6}
                    &\multicolumn{1}{c}{(1)}&\multicolumn{1}{c}{(2)}&\multicolumn{1}{c}{(3)}&\multicolumn{1}{c}{(4)}&\multicolumn{1}{c}{(5)}\\
                    &\multicolumn{1}{c}{Value}&\multicolumn{1}{c}{Value}&\multicolumn{1}{c}{Value}&\multicolumn{1}{c}{Quantity}&\multicolumn{1}{c}{Price}\\
\hline
Tariff x Post       &      -0.273       &      -0.255       &      -0.553\sym{a}&      -0.551\sym{a}&      -0.002       \\
                    &     (0.196)       &     (0.196)       &     (0.132)       &     (0.105)       &     (0.062)       \\
\hline
Observations        &        4240       &        2265       &        2265       &        2265       &        2265       \\
Pseudo-R2           &       0.920       &       0.955       &           .       &           .       &           .       \\
Adj.-R2             &           .       &           .       &       0.825       &       0.801       &       0.618       \\
\hline\hline
\end{tabular}
}

		\caption*{\begin{footnotesize}Note: Regressions estimated with monthly export data from January 2017 to February 2021. The dependent variables in columns~1, 2, 3, and 4 are import revenue, log import revenue, log import quantity, and log import price, respectively. All regressions include product$\times$destination, year, and month fixed effects. Tariff$\times$Post turns 1 if a product was subject to a tariff in a specific import origin and the month-year combination is within the tariff-imposition period. Standard errors clustered at the product level are in parentheses. a represents statistical significance at the 1\%.
		\end{footnotesize}}
	\end{center}
\end{table}

In columns~3 to~5, I decompose the change in the value of US imports into the quantity and price margins. Since this decomposition can only be performed with an OLS estimator, in column~2 I repeat the PPML estimation in column~1 with the sample used for the OLS decomposition. The point estimate barely changes relative to the estimate reported in column~1. However, when an OLS model is used to estimate Equation~\eqref{eq:olives}, the interaction coefficient becomes negative and significant (column~3). The table also shows that the reduction in revenue was explained by a reduction in quantities. Prices, which do not include tariff duties, did not change, indicating a complete pass-through of tariffs to prices. This result is in line with the low import demand elasticity of Spanish olives in the US \citep{fajgelbaum2022uschina}.

\section{Conclusion}
\label{sec:conclusion}

In contrast to previous decades characterized by a process of trade liberalization at the global and regional level, there has been a shift towards more unilateral and aggressive trade policies since the middle 2010s. This paper has analyzed how exporters behave in this hostile trade environment.

I use the retaliatory tariffs that the US imposed on some EU products between October 2019 and March 2021 due to the Airbus-Boeing trade dispute as a case study. I show that US imports of Spanish products targeted by tariffs declined by 62\% relative to other US imports from Spain. However, the export revenue of Spanish exporters of tariff-targeted products did not significantly decrease relative to the one of other Spanish exporters to the US during the period in which tariffs were in force. I show that Spanish exporters used two trade-avoidance strategies to mute the negative effects of tariffs. Firstly, instead of exporting Spanish products targeted by tariffs, they exported the same products but originated in countries not targeted by tariffs. Secondly, they shifted from varieties targeted by tariffs to varieties that had not been targeted by tariffs within the same product category.

The trade avoidance strategy followed by Spanish exporters to the US enabled them to neutralize the negative effects of tariffs. This shows that exporters, at least those specialized in primary and manufactures of primary products, have an additional margin to counteract a tariff hike.



\begin{thebibliography}{}
	
	\bibitem[Albornoz et~al., 2021]{albornoz2021tariffhikes}
	Albornoz, F., Brambilla, I., and Ornellas, E. (2021).
	\newblock Firm export responses to tariff hikes.
	\newblock {\em Centre for Economic Performance Working Paper 1683}.
	
	\bibitem[Alessandria et~al., 2019]{alessandria2019taking}
	Alessandria, G.~A., Khan, S.~Y., and Khederlarian, A. (2019).
	\newblock Taking stock of trade policy uncertainty: evidence from {China's}
	{Pre-WTO} accession.
	\newblock {\em NBER Working Paper 25965}.
	
	\bibitem[Amiti et~al., 2019]{amiti2019impact}
	Amiti, M., Redding, S.~J., and Weinstein, D.~E. (2019).
	\newblock The impact of the 2018 tariffs on prices and welfare.
	\newblock {\em Journal of Economic Perspectives}, 33(4):187--210.
	
	\bibitem[Atkeson and Burstein, 2008]{atkeson2008pricing}
	Atkeson, A. and Burstein, A. (2008).
	\newblock Pricing-to-market, trade costs, and international relative prices.
	\newblock {\em American Economic Review}, 98(5):1998--2031.
	
	\bibitem[Bernard et~al., 2011]{bernard2011multiproduct}
	Bernard, A.~B., Redding, S.~J., and Schott, P.~K. (2011).
	\newblock Multiproduct firms and trade liberalization.
	\newblock {\em The Quarterly Journal of Economics}, 126(3):1271--1318.
	
	\bibitem[Bhagwati, 1964]{bhagwati1964underinvoicing}
	Bhagwati, J. (1964).
	\newblock On the underinvoicing of imports.
	\newblock {\em Bulletin of the Oxford University Institute of Economics and
		Statistics}, 29:61--77.
	
	\bibitem[Cavallo et~al., 2021]{cavallo2021passthrough}
	Cavallo, A., Gopinath, G., Neiman, B., and Tang, J. (2021).
	\newblock Tariff pass-through at the border and at the store: Evidence from
	{US} trade policy.
	\newblock {\em American Economic Review: Insights}, 3(1):19--34.
	
	\bibitem[Colantone et~al., 2021]{colantone2021backlash}
	Colantone, I., Ottaviano, G.~I., and Stanig, P. (2021).
	\newblock The backlash of globalization.
	\newblock {\em Centre for Economic Performance Working Paper 1800}.
	
	\bibitem[Demir and Javorcik, 2020]{demir2020benfordslaw}
	Demir, B. and Javorcik, B. (2020).
	\newblock Trade policy changes, tax evasion and {B}enford's law.
	\newblock {\em Journal of Development Economics}, 144:102456.
	
	\bibitem[Dhingra and Sampson, 2022]{dhingra2022brexit}
	Dhingra, S. and Sampson, T. (2022).
	\newblock Expecting {B}rexit.
	\newblock {\em Annual Review of Economics}, 14(1):495--519.
	
	\bibitem[Eckel and Neary, 2010]{eckel2010multi}
	Eckel, C. and Neary, J.~P. (2010).
	\newblock Multi-product firms and flexible manufacturing in the global economy.
	\newblock {\em The Review of Economic Studies}, 77(1):188--217.
	
	\bibitem[Fajgelbaum et~al., 2020]{fajgelbaum2020return}
	Fajgelbaum, P.~D., Goldberg, P.~K., Kennedy, P.~J., and Khandelwal, A.~K.
	(2020).
	\newblock The return to protectionism.
	\newblock {\em The Quarterly Journal of Economics}, 135(1):1--55.
	
	\bibitem[Fajgelbaum and Khandelwal, 2022]{fajgelbaum2022uschina}
	Fajgelbaum, P.~D. and Khandelwal, A.~K. (2022).
	\newblock The economic impacts of the {US-China} trade war.
	\newblock {\em Annual Review of Economics}, 14(1):205--28.
	
	\bibitem[Fisman and Wei, 2004]{fisman2004tax}
	Fisman, R. and Wei, S.~J. (2004).
	\newblock Tax rates and tax evasion: evidence from ``missing imports'' in
	{C}hina.
	\newblock {\em Journal of Political Economy}, 112(2):471--496.
	
	\bibitem[Fitzgerald and Haller, 2018]{fitzgerald2018shocks}
	Fitzgerald, D. and Haller, S. (2018).
	\newblock Exporters and shocks.
	\newblock {\em Journal of International Economics}, 113(July):154--171.
	
	\bibitem[Flaaen et~al., 2020]{flaaen2020production}
	Flaaen, A., Horta{\c{c}}su, A., and Tintelnot, F. (2020).
	\newblock The production relocation and price effects of {US} trade policy: the
	case of washing machines.
	\newblock {\em American Economic Review}, 110(7):2103--27.
	
	\bibitem[Iacovone et~al., 2015]{iacovone2015walmart}
	Iacovone, L., Javorcik, B., Keller, W., and Tybout, J. (2015).
	\newblock Supplier responses to {W}almart's invasion in {M}exico.
	\newblock {\em Journal of International Economics}, 95(1):1--15.
	
	\bibitem[Javorcik and Narciso, 2008]{javorcik2008differentiated}
	Javorcik, B.~S. and Narciso, G. (2008).
	\newblock Differentiated products and evasion of import tariffs.
	\newblock {\em Journal of International Economics}, 76(2):208--222.
	
	\bibitem[Jiao et~al., 2022]{jiao2022impacts}
	Jiao, Y., Liu, Z., Tian, Z., and Wang, X. (2022).
	\newblock The impacts of the {US} trade war on {Chinese} exporters.
	\newblock {\em Review of Economics and Statistics}.
	\newblock Conditionally accepted.
	
	\bibitem[Khan and Khederlarian, 2021]{khan2021does}
	Khan, S.~Y. and Khederlarian, A. (2021).
	\newblock How does trade respond to anticipated tariff changes? {E}vidence from
	{NAFTA}.
	\newblock {\em Journal of International Economics}, 133:103538.
	
	\bibitem[Ludema and Yu, 2016]{ludema2016passthrough}
	Ludema, R.~D. and Yu, Z. (2016).
	\newblock Tariff pass-through, firm heterogeneity and product quality.
	\newblock {\em Journal of International Economics}, 103:234--249.
	
	\bibitem[Melitz, 2003]{melitz2003impact}
	Melitz, M.~J. (2003).
	\newblock The impact of trade on intra-industry reallocations and aggregate
	industry productivity.
	\newblock {\em Econometrica}, 71(6):1695--1725.
	
	\bibitem[Santos-Silva and Tenreyro, 2010]{santossilva2010ppml}
	Santos-Silva, J. and Tenreyro, S. (2010).
	\newblock Further simulation evidence on the performance of the {P}oisson
	pseudo-maximum likelihood estimator.
	\newblock {\em Economics Letters}, 112(2):220 -- 222.
	
\end{thebibliography}

\clearpage

\appendix \label{app:all}

\setcounter{equation}{0}
\renewcommand\theequation{A.\arabic{equation}}


\setcounter{figure}{0}
\renewcommand\thefigure{A.\arabic{figure}}

\setcounter{table}{0}
\renewcommand\thetable{A.\arabic{table}}


\beginappendix
\begin{appendices}
	
	\begin{table}[htbp]
		\begin{center}
			\caption{Triple-difference. Impact of tariffs on US imports from Spain. Product-level analysis}
			\label{tab:product_triple}
			{
\def\sym#1{\ifmmode^{#1}\else\(^{#1}\)\fi}
\begin{tabular}{l*{3}{c}}
\hline\hline
                    &\multicolumn{1}{c}{(1)}&\multicolumn{1}{c}{(2)}&\multicolumn{1}{c}{(3)}\\
                    &\multicolumn{1}{c}{Value}&\multicolumn{1}{c}{Quantity}&\multicolumn{1}{c}{Price}\\
\hline
Tariff x Post       &      -0.578\sym{a}&      -0.558\sym{a}&      -0.020       \\
                    &     (0.147)       &     (0.151)       &     (0.046)       \\
\hline
Observations        &     3065942       &     3065942       &     3065942       \\
Adj.-R2             &       0.669       &       0.760       &       0.823       \\
\hline\hline
\end{tabular}
}

			\caption*{\begin{footnotesize}Note: Regressions estimated with monthly export data from January 2017 to February 2021. The dependent variables in column~1,~2, and~3 are log import revenue, log import quantity, and log import price, respectively. All regressions include product$\times$destination, year$\times$destination, month$\times$destination, product$\times$year, and product$\times$month fixed effects. Tariff$\times$Post turns 1 if a product was subject to a tariff and the month-year combination is within the tariff-imposition period. Standard errors clustered at the product level are in parentheses. a represents statistical significance at the 1\% level.
			\end{footnotesize}}
		\end{center}
	\end{table}

	\begin{table}[htbp]
		\begin{center}
			\footnotesize
			\caption{Top-4 product categories. Impact of tariffs on US exports}
			\label{tab:int_categories}
			{
\def\sym#1{\ifmmode^{#1}\else\(^{#1}\)\fi}
\begin{tabular}{l*{4}{c}}
\hline\hline
                    &\multicolumn{1}{c}{(1)}&\multicolumn{1}{c}{(2)}&\multicolumn{1}{c}{(3)}&\multicolumn{1}{c}{(4)}\\
                    &\multicolumn{1}{c}{Olive oil}&\multicolumn{1}{c}{Wine}&\multicolumn{1}{c}{Olives}&\multicolumn{1}{c}{Cheese}\\
\hline
Share tariff x Tariff period&      -0.096       &      -0.118       &       0.112       &      -0.049       \\
                    &     (0.195)       &     (0.096)       &     (0.255)       &     (0.403)       \\
\hline
Observations        &       11048       &       86353       &        3721       &        4313       \\
Pseudo-R2           &       0.860       &       0.771       &       0.868       &       0.853       \\
\hline\hline
\end{tabular}
}

			\caption*{\begin{footnotesize}Note: The dependent variable is monthly exports to the US. All regressions include firm, year$\times$month fixed effects, and year-specific dummies to control for firm size. Standard errors clustered at the firm level are in parentheses.
			\end{footnotesize}}
		\end{center}
	\end{table}

\begin{table}[htbp]
	\begin{center}
		\caption{Correlation between the increase in imports of olive oil from Tunisia and Portugal and exports to the US}
		\label{tab:simple_ols}
		{
\def\sym#1{\ifmmode^{#1}\else\(^{#1}\)\fi}
\begin{tabular}{l*{1}{c}}
\hline\hline
                    &\multicolumn{1}{c}{(1)}       \\
\hline
$\Delta$ Imports from Portugal and Tunisia&       0.627\sym{b}\\
                    &     (0.251)       \\
[1em]
Constant            &    2360.698\sym{a}\\
                    &   (829.509)       \\
\hline
Observations        &         198       \\
\(R^{2}\)           &       0.031       \\
\hline\hline
\end{tabular}
}

		\caption*{\begin{footnotesize}Note: The dependent variable is the value of olive oil exports to the US during the tariff period. Standard errors are in parentheses.
				a and b denote statistical significance at the 1 and 5\% levels, respectively.
		\end{footnotesize}}
	\end{center}
\end{table}

\end{appendices}
\end{document}